\documentclass{aa}
\usepackage{amsmath}
\usepackage{graphicx}
\usepackage{natbib}
\usepackage{txfonts}
\bibpunct{(}{)}{;}{a}{}{,}
\newcommand{\Ds}{D_{\mathrm{s}}}

\newcommand{\Dls}{D_{\mathrm{ls}}}

\newcommand{\myarcsec}{\hbox{$.\!\!^{\prime\prime}$}}
\newcommand{\myarcmin}{\hbox{$.\!\!^{\prime}$}}
\newcommand{\myarcsecnodot}{\hbox{$\;\!\!^{\prime\prime}\;$}}
\newcommand{\myarcminnodot}{\hbox{$^{\prime}\;$}}

\def\cm3{\rm ~cm^{-3}}
\def\kms{\rm ~km~s^{-1}}

\begin{document}
 
\bibliographystyle{aa}

\title{Mass, Light and Colour of the Cosmic Web \\
  in the Supercluster SCL2243-0935 (z=0.447) 
\thanks{This work is based on observations obtained 
  with MegaPrime/MegaCam, a joint project of CFHT and
  CEA/DAPNIA, at the Canada-France-Hawaii Telescope (CFHT) which
  is operated by the National Research Council (NRC) of Canada,
  the Institut National des Sciences de l'Univers of the Centre
  National de la Recherche Scientifique (CNRS) of France, and the
  University of Hawaii (programme ID: 2008BO01); based on 
  observations made with ESO Telescopes at the La Silla and Paranal 
  Observatories, Chile (ESO Programmes 165.S-0187 and 079.A-0063); based on
  observations made with the 2.5m Isaac Newton Telescope operated 
  on the island of La Palma by the Isaac Newton Group in the Spanish 
  Observatorio del Roque de los Muchachos of the Instituto de 
  Astrofísica de Canarias (programme ID 2008B/C11 and 2009B/C1).}}

\author{Mischa Schirmer\inst{1,2}
  \and
  Hendrik Hildebrandt\inst{3,4}
  \and
  Konrad Kuijken\inst{3}
  \and
  Thomas Erben\inst{1}
}

\offprints{mischa@astro.uni-bonn.de}

\institute{Argelander-Institut f\"ur Astronomie, Universit\"at
  Bonn, Auf dem H\"ugel 71, 53121 Bonn, Germany
  \and
  Isaac Newton Group of Telescopes, Calle Alvarez Abreu 70, 38700
  Santa Cruz de La Palma, Spain
  \and
  Leiden Observatory, Leiden University, Niels Bohrweg 2, 2333 Leiden, 
  The Netherlands
  \and
  Department of Physics and Astronomy, 6224 Agricultural Road,
  University of British Columbia, Vancouver, Canada
}

\date{Received; accepted}

\abstract 
{}
{In archival 2.2m MPG-ESO/WFI data we discovered several mass peaks
  through weak gravitational lensing, forming a possible supercluster
  at redshift $0.45$. Through wide-field imaging and spectroscopy we 
  aim to identify the supercluster centre, confirm individual 
  member clusters, and detect possible connecting filaments.}
{Through multi-colour imaging with CFHT/Megaprime and INT/WFC we 
  identify a population of early-type galaxies and use it to trace
  the supercluster network. EMMI/NTT multi-object spectroscopy is used
  to verify the initial shear-selected cluster candidates. We use weak
  gravitational lensing to obtain mass estimates for the supercluster
  centre and the filaments.}
{We identified the centre of the SCL2243-0935 supercluster, MACS
  J2243-0935, which was found independently by Ebeling et 
  al. (2001,2010). We found 13 more clusters or
  overdensities embedded in a large filamentary network. Spectroscopic
  confirmation for about half of them is still pending. Three 
  $(5-15)\,h_{70}^{-1}$ Mpc filaments are detected, and we estimate
  the global size of SCL2243 to be $45\times15\times50\,h_{70}^{-1}$
  Mpc, making it one of the largest superclusters known at intermediate
  redshifts. Weak lensing yields
  $r_{200}=(2.06\pm0.13)\,h_{70}^{-1}$ Mpc and 
  $M_{200}=(1.54\pm0.29)\times 10^{15}\,M_\odot$ for MACS J2243 with
  $M/L=428\pm82$, very similar to results from size-richness cluster 
  scaling relations. Integrating the weak lensing surface mass density
  over the supercluster network (defined by increased $i$-band
  luminosity or $g-i$ colours), we find
  $(1.53\pm1.01)\times10^{15}\,M_\odot$ and $M/L=305\pm201$ for the 
  three main filaments, 
  consistant with theoretical predictions. The filaments' projected 
  dimensionless surface mass density $\kappa$ varies between $0.007-0.012$,
  corresponding to $\rho/\rho_{\rm crit}=10-100$ depending on
  location and de-projection. The greatly varying density of the 
  cosmic web is also reflected in the mean colour of 
  galaxies, e.g. $\langle g-i\rangle=2.27$ mag for the
  supercluster centre and $1.80$ mag for the filaments.}
{SCL2243 is significantly larger and much more richly structured
  than other known superclusters such as A901/902 or MS0302 studied
  with weak lensing before. It is a text-book supercluster with little
  contamination along the line of sight, making it a perfect sandbox
  for testing new techniques probing the cosmic web.}


\titlerunning{Mass, Light and Colour of the Cosmic Web in SCL 2243-0935 ($z=0.45$)}

\authorrunning{Schirmer et al.}

\maketitle

\section{\label{intro}Introduction}
Superclusters of galaxies mark the largest and most massive 
structures known in the Universe, tracing the transition
from the linear to the non-linear regime in the evolution of 
the density contrast. Being non-virialised, they usually come 
in the form of long filaments \citep{bkp96}, and it is not a priori 
clear that they form gravitationally bound systems \citep{grs02}. 
The evolution of a supercluster's global structure is driven by 
gravitation only, as hydrodynamic or intergalactic processes become 
important on cluster scales and below. Beyond a certain extent
superclusters are also very susceptible to (accelerated) 
cosmic expansion. \cite{arm09} have shown that in the $\Lambda CDM$
paradigm superclusters can withstand expansion and will eventually 
collapse, provided that their average density inside a spherical shell
exceeds $2.36\,\rho_{\rm{crit}}$.

\begin{table*}[t]
  \caption{\label{knownsuperclusters} Known superclusters at redshifts
    $0.2<z<1.0$ with at least three member clusters. In addition to
    redshift and the number of member clusters (or sub-structures in
    filaments), we report the extent transverse and along the line of
    sight, and an estimate for the total mass. Also included are A1437
    at lower redshift, as it is similar in size and richness to
    SCL2243-0935, and A901/2 because together with MS0302 they are the
    only other two superclusters studied with weak lensing.}
  \begin{tabular}{lccccccl}
    \noalign{\smallskip}
    \noalign{\smallskip}
    \hline 
    \hline 
    \noalign{\smallskip}
    \noalign{\smallskip}
    Name & RA & DEC & $z$ & N(cluster) & $r_{\rm trans}\times r_{\rm los}\;[h^{-1}$Mpc] & $M_{\rm{tot}}$
    [$\times10^{14} M_\odot$] & Further reading\\ 
    \noalign{\smallskip}
    \noalign{\smallskip}
    \hline 
    \noalign{\smallskip}
    CL0016     & 00:18:30 & 16:25:00  & $0.542-0.550$ & 5    & $15\times15$  & $30-59$ & \cite{thk07}\\
    MS0302     & 03:05:26 & 17:17:54  & 0.423         & 3    & $6\times20$   & $6.6\pm1.0$ &\cite{gmf04}\\
    A901/902   & 09:56:12 & -09:58:12 & 0.165         & 4    & 3.0           & $7.24\pm1.05$ & \cite{hgp08}\\  
    J1000+0231 & 10:00:12 & 02:31:12  & $0.65-0.85$   & 3    & $13\times400$ & $60\pm30$ & \cite{gcf07}\\
    A1437      & 12:00:28 & 03:20:18  & $0.125-0.138$ & 7    & $87\times46$  & $-$ & \cite{wou09}\\
    CL1604     & 16:04:14 & 43:15:00  & $0.84-0.96$   & 9    & $13\times100$ & $>3.6$ & \cite{gll08}\\
    DXS1609    & 16:09:00 & 54:30:00  & $0.88-0.90$   & 5    & $28\times28$  & $>5$ & \cite{swinbank07}\\
    SCL2243-0935 & 22:43:00 & -09:35:00 & $0.435-0.456$ & $>5$ & $45\times50$  & $30\pm15 $ & This paper\\  
    \hline
  \end{tabular}
  \normalsize
\end{table*}

The morphologies and appearances of superclusters are quite 
diverse. \cite{ees07} divided their sample of 543 superclusters 
selected from the 2dF Galaxy Redshift Survey
\citep[][]{colless01} into four richness classes, parametrised
by the number of peaks in the smoothed and luminosity-weighted density
field. Poor superclusters contain $1-2$ such clusters, whereas
richer systems have several large compact cores of high density
and show a greater morphological diversity \citep[][]{esl07}.
Furthermore, supercluster cores with masses higher than
$10^{15}\,M_\odot$ are more entangled in the cosmic web, having
about five connected filaments as compared to ten times
less massive clusters \citep[on average two filaments][]{awj10}.

Galaxies in superclusters have been well investigated in
particular for low redshift systems 
\citep[e.g.][]{mmh10,eet07,ets10}. This is facilitated by 
the fact that these superclusters decompose into many individual
Abell-type clusters which have already been studied in
detail. Galaxies living in the inter-cluster environment 
also get attention. For example, \cite{prp08} found increased star
formation rates for galaxies located in filaments well beyond the
virial cluster radius, likely triggered by close encounters in the
enhanced density field. \cite{pqc06} focused on more global
properties, showing that nearly half of the total galaxy population in
the Shapley supercluster is located outside clusters and contributes
up to twice as much mass as cluster member galaxies.

On global scales superclusters are rarely investigated, in particular
those with $z>0.2$ and especially when it comes to the relation
between luminous and dark matter. The latter has been well probed for
individual clusters, mostly based on weak gravitational lensing and
X-ray imaging  
\citep[see e.g.][]{cbg06,seh07,jsw07,rmb08,sjm09}. Superclusters,
however, are particularly difficult to study in this respect. First,
with sizes of $10-100$ Mpc they span many degrees on sky when at low
redshift, much larger than the fields attainable with current
telescopes. Both weak lensing and X-ray observations require long
exposure times, hence such surveys would be extremely
expensive. Second, the strength of the weak lensing effect scales with
$\Dls/\Ds$, the ratio of the angular diameter distances between the
lens and the source, and between the observer and the source. Thus
lensing is inefficient for clusters with $z<0.1$, and difficult for
clusters with $z\gtrsim0.7$ as the number density of lensed background
sources is greatly diminished. Weak lensing studies of superclusters
are therefore limited to targets extending not much more than one
degree on sky, limiting the redshift range to $0.2-0.7$.

A list of the handful of superclusters studied over the last 15 years
in this redshift range is shown in Table \ref{knownsuperclusters}.
We ignore compact and more common double
clusters, requiring a minimum number of three distinct
member clusters\footnote{These are all objects found searching 
for the term \textit{supercluster} in publications
between 1995 and 2010, and in referenced papers}. About half of
these superclusters were serendipitous discoveries in deep survey
fields. We also include two examples with redshifts less than $0.2$:
A1437, as it appears to be comparable in size and richness to
SCL2243-0935 (the cluster we study in this paper, hereafter SCL2243), 
and A901/902 as it was already the subject of two weak lensing
campaigns. Even if our literature survey is incomplete, it shows that
only few such superclusters are known at intermediate
redshifts. Deeper spectroscopic surveys will certainly change that, as
can be extrapolated from \cite{ees07} who identified 543 superclusters
with $z<0.2$.

To date, MS0302 \citep[][]{kwl98,gmf04} and A901/2 
\citep[][]{gtm02,hgp08,dgh10} are the only two superclusters that have
been studied with weak lensing. With an extent of only $\sim3$ Mpc, 
A901/902 is a very compact system, whereas MS0302 is more characteristic
for a supercluster with $6-20$ Mpc. Both are low mass systems with
$M_{\rm tot}\sim(6-7)\times10^{14}M_\odot$, compared to several
$10^{15}M_\odot$ or even $10^{16}M_\odot$ for the most massive
structures known \citep[e.g.][for the Shapley supercluster]{pqc06}.

In this paper we present our analysis of the supercluster 
SCL2243. We discovered it in our 19 square degree weak lensing 
survey conducted with the wide field imager at the 2.2m MPG-ESO
telescope, aiming for the blind detection of galaxy clusters 
\citep{seh07}. Several mass peaks were identified at redshift
$z\sim0.44$. Using multi-colour follow-up observations with
CFHT/Megaprime and INT/WFC we covered an area of four square degrees
around the initial detections. Our objectives are the
identification of a possible supercluster centre and of further member
clusters, and the spectroscopic verification of three of the initially
shear-selected cluster candidates in the MPG-ESO/WFI data. SCL2243 is
the first supercluster that was discovered using the shear-selection
technique. The supercluster core, MACS 2243-0935, was found
independently by \cite{eeh01} through X-ray selection. Out of 
the three superclusters analysed with weak lensing, SCL2243 is by far
the most massive one, and also the one with the highest redshift.

This paper is organised as follows: In Sect. 2 we describe the
observations and the data reduction. In Sect. 3 we select early-type
galaxies as tracers for the underlying supercluster skeleton and
identify possible member clusters and filaments.
Using weak lensing, we reconstruct the dark matter distribution and
obtain mass-to-light ratios. We discuss our results in Sect. 4 and
summarise in Sect. 5. 

Throughout this paper we assume a flat standard
cosmology with $\Omega_m=0.27$, $\Omega_\Lambda=0.73$ and
$H_0=70\,h_{70}\kms\,{\rm Mpc}^{-1}$. On occasion we refer to
relations from the literature with 
$H_0=100\,h_{100}\kms\,{\rm Mpc}^{-1}$. Optical  
luminosities are given in solar units for the $i$-band rest-frame, using
$M_{i,\odot}=4.52$ mag. The relation between physical and angular scales
at $z=0.45$ is $1.0\,h_{70}^{-1} {\rm Mpc} = 2\myarcmin87$. Error bars
represent the $1\sigma$ confidence level unless stated otherwise.

Note that in N-body simulations, the term filament is used to
describe the low-density diffuse dark matter component connecting
one or several clusters. In a filament, more or less distinct 
groups and smaller clusters of galaxies can form, tracing the
underlying dark matter filament. For simplicity, in this
paper we also refer to these optical structures as filaments.

\section{Observations and data reduction}
All images were processed with THELI\footnote{THELI is freely
available at http://www.astro.uni-bonn.de/$\sim$theli}
\citep{esd05}. The instrumental signature was removed through 
overscan correction, debiasing, flat-fielding and, if necessary, 
defringing. After fixing astrometry and photometry, the sky was
subtracted and the data combined into mosaics. Particularities in
the data reduction are explained below. The various
pointings on sky are shown in Fig. \ref{fields}, and their
characteristics are summarised in Table \ref{obstable}.

\subsection{2.2m MPG-ESO/WFI imaging}
The initial shear-selection discovery of SCL2243 was done with archival
$R$-band data taken 2000-08-28/29 in clear and good seeing conditions 
with the Wide Field Imager (WFI) at the 2.2m MPG-ESO telescope in La 
Silla, Chile. The objective of the programme (165.S-0187) was a TNO 
search; no TNO was found in that field (O. Hainaut, priv. comm.). 
The data were identified by us in the ESO archive using the
\textit{Querator}\footnote{http://archive.eso.org/querator} tool. It 
was developed in the course of our ASTROVIRTEL proposal, allowing for
a more specific data mining of the ESO archive returning fields
suitable for weak lensing purposes (in this case identifying data sets 
with sub-arcsec seeing conditions and a certain minimum total
integration time in $R$-band). More details can be found in
\cite{seh07} and \cite{mpb04}. WFI has a field of view of 
34\myarcminnodot$\times$34\myarcminnodot with a pixel scale of 
0\myarcsec238. 

An absolute photometric calibration was not done as it
was not necessary for our initial single-band weak lensing 
survey. We simply adopted the standard instrumental ZP for $R$-band,
yielding a $5\sigma$ depth of $R_{\rm lim}\sim24.7$ mag.

\subsection{\label{photz}CFHT/Megaprime imaging and photometric redshifts}
\begin{figure}[t]
  \includegraphics[width=1.0\hsize]{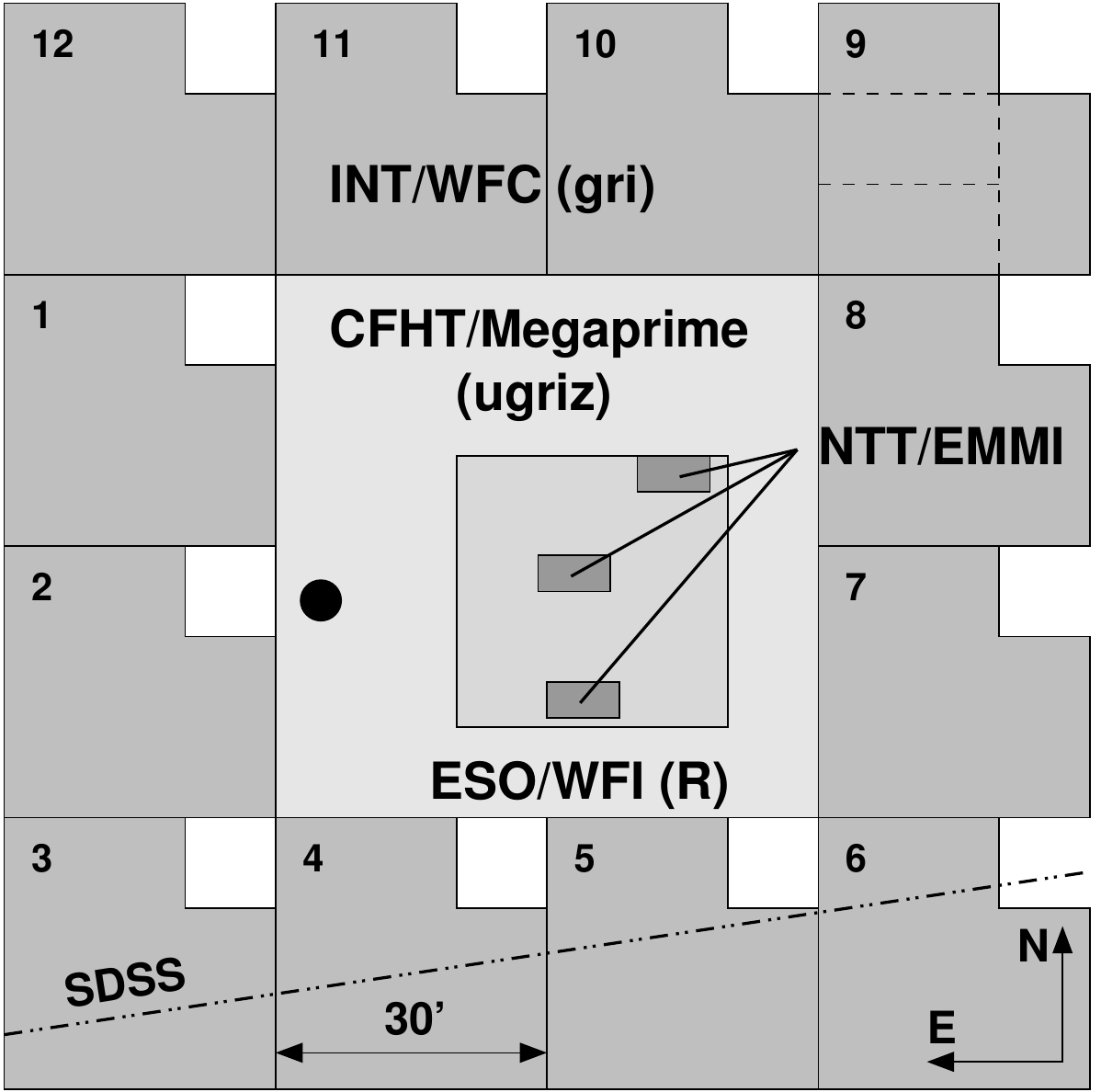}
  \caption{\label{fields}{Pointings of the INT, CFHT and ESO imaging 
      data sets together with the filters used, and the three EMMI
      spectroscopic fields. The position of MACS J2243, the
      supercluster centre, is indicated by the black dot. It was 
      unknown by the time the quasi-simultaneous CFHT and INT
      observations were performed. The layout of the 4 CCDs in
      the WFC/INT detector array is indicated for the upper right
      pointing. The area above the dash-dotted line is covered by 
      SDSS-DR7.}}
\end{figure}

With the moderate depth of the single-band WFI data and only a crude
magnitude cut for the selection of lensed background galaxies, our weak
lensing analysis of clusters at $z\sim0.45$ was limited. To cover
a larger area and to refine the weak lensing measurements, we obtained 
CFHT/Megaprime $ugriz$ images in excellent and photometric seeing
conditions through Opticon proposal 2008BO01, with a pixel scale of 
0\myarcsec186. The pointing for CFHT was chosen to fully cover the WFI 
field (see Fig. \ref{fields}). We did not centre it on the WFI
pointing to keep bright stars outside the field of view.

Images were pre-reduced using {\tt ELIXIR} \citep{mac04} at CFHT,
including corrections for scattered light on the order of 0.1 mag. The
remaining processing was done with THELI following \cite{ehl09}. 
Astrometry and relative photometry was performed using Scamp 
\citep{ber06}, processing all filters at the same time and thus
guaranteeing a common distortion correction and pixel grid for 
all coadded images. For the absolute flux calibration we adopted the
standard {\tt ELIXIR} zeropoints as data were taken in photometric 
conditions. Zeropoints were refined during the photometric redshift 
calibration (see Sect. \ref{photz2}).

\subsubsection{\label{asteroid}Asteroid masking: an outlier rejection filter for SWarp}
SCL2243 is projected near the ecliptic, resulting in a large number of
asteroid tracks in the images. This is problematic as the fainter ones
with low proper motion can be mistaken for highly elongated 
(i.e. significantly sheared) background galaxies. For the
single-band MPG-ESO/WFI data asteroids were removed manually from 
the object catalogue obtained from the coadded image. Manual cleaning
became unfeasible for our multi-colour CFHT and INT observations with
16 times larger sky coverage. Instead of working on the catalogue
level, we developed an advanced outlier rejection filter which 
operates on the individual resampled images before image
combination. In this way a coadded image free from asteroid tracks is
created, providing a much cleaner basis to work with.
As this tool is a recent implementation in THELI and not described in 
\cite{esd05}, we provide more details in the following.

The working principle of the filter is as follows. THELI uses 
{\tt SWarp} \citep{ber02} to resample images to a common grid on sky
and to perform the final image coaddition. During the resampling
process all astrometric corrections and projections are applied,
meaning that a particular pixel in one exposure covers a well-defined 
area on sky. Precisely the same area is covered by the other
(resampled) pixels in subsequent exposures. The standard procedure in
{\tt SWarp} is to calculate a weighted mean from the
resampled pixels to estimate the flux in the coadded image,
\begin{equation}
I_{\rm coadd}=\frac{\sum_i I_i f_i w_i}{\sum_i w_i}\,.
\end{equation}
Here $I_i$ is the flux of the resampled pixel in the $i$-th exposure, 
$f_i$ a flux correction factor (containing exposure times and relative 
photometric zeropoints), and $w_i$ an individual weight calculated
from the normalised flat, sky noise and $f_i$ 
\citep[see Sect. 7 in][for details]{esd05}. Whereas all the
information is at hand, {\tt SWarp} (as of version 2.19) has no
internal outlier rejection implemented. However, we can work around
this by identifying bad pixels externally and setting them to zero 
in the weight maps before {\tt Swarp} performs the weighted image 
combination. In this manner we do not interfere with the internal
processing of {\tt Swarp}, automatically preserve photometry and do
not introduce biases.

\begin{figure}[t]
  \includegraphics[width=1.0\hsize]{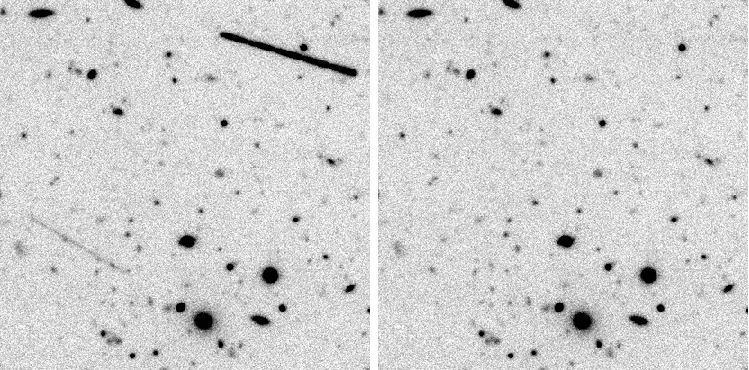}
  \caption{\label{asteroidfig}{Left: Coadded CFHT $r$-band image without
      asteroid filtering. Right: Using the rejection algorithm described
      in Sect. \ref{asteroid}.}}
\end{figure}

To identify the bad pixels, we reconstruct the pixel stacks from the
resampled images. Ignoring the highest and lowest pixel values in the
stack, we calculate the standard deviation and, in a first pass, 
apply a classic $\kappa-\sigma$-clipping to identify bad pixels and
asteroids. To mask fainter asteroids we have to use comparatively low 
values of $\kappa\sim 3-4$. This results in zealous masking of healthy
pixels in the presence of non-Gaussian (Poissonian) noise and small 
stack sizes. However, since the PSF is well sampled and since the
images were resampled by a broad Lanczos3 kernel, an asteroid causing
a bad pixel will also spoil the neighbouring pixels. Therefore, 
in a second pass, THELI only masks a bad pixel if at least $n$ of the 
8 surrounding pixels were initially also recognised as bad ($n$ is a 
user-supplied parameter and we find that $n=4-6$ works well). With
this clustering analysis we identify and mask asteroids only while 
preserving the noise properties of the rest of the image. Optionally, 
pixels adjacent to a cluster of bad pixels can also be masked, which 
helps removing the haloes of very bright asteroids. See
Fig. \ref{asteroidfig} for an example.

\subsubsection{\label{photz2}Photometric redshifts}
Photometric redshifts were obtained as outlined in \cite{hpe09}, with
some modifications as outlined in the following. Prior to the
photo-z determination, the PSF in all filters was homogenised and also
brought to a Gaussian shape. This method will be described in a
forthcoming paper (Kuijken et al., in prep.). Our experience with
CFHTLS \citep{ehl09} and deeper spectra showed that this 
improves in particular the photo-z estimates of faint and high
redshift galaxies. The CFHT data used for the present paper are very
similar to those of CFHTLS, hence it is plausible to assume that they
profited from the PSF homogenisation in a similar way, even though we
cannot verify it due to the lack of deep spectroscopic coverage. 

Using 89 SDSS \citep{aaa08} and 101 NTT/EMMI spectra, we re-calibrated
the photometric zeropoints of the CFHT/Megaprime images. Only sources
with photometric errors smaller than 0.1 mag in all bands were used
for this purpose. In detail, we fix the redshifts of the corresponding
galaxies to their spectroscopically determined values. The average
magnitude differences between the best-fit templates and the observed
photometry are zero within errors, with exception of
$g$-band ($0.047$ mag, demonstrating the excellent photometric
calibration using {\tt ELIXIR}). The photo-z uncertainty is
$\sigma_{\Delta z}\sim0.04\,(1+z)^{-1}$, identical to the one for our
reduction of the CFHTLS fields \citep{ehl09}.

\subsection{\label{intimaging}INT/WFC imaging}
To probe the outer regions of SCL2243 we observed 12 adjacent
pointings in $gri$ filters using the wide field camera (WFC) at the
2.5m Isaac Newton Telescope (INT) in La Palma, Spain. INT/WFC has four
detectors forming an asymmetric focal plane (see upper right of
Fig. \ref{fields}), covering $34^{\prime}\times34^{\prime}$ on sky
with a pixel scale of 0\myarcsec331. INT/WFC cannot be used for weak
lensing due to its large pixels and image quality. The latter is
degraded by dome seeing, a non-coplanar detector
arrangement, and sometimes by tracking instabilities. Sub-arcsecond
image seeing is infrequently achieved with INT/WFC. However, it 
is a good photometric instrument with high throughput and thus
well suited for our survey of SCL2243.

Data were taken mostly in photometric conditions in 2008-10-24 to 
2008-10-27, 2008-11-16 and in 2009-10-14 to 2009-10-16 through CAT-ID
2008B/C11 and 2009B/C01. Images are shallower than
the CFHT data (see Table \ref{obstable}) due to shorter exposure 
times and worse seeing conditions. Pointings 1 and 2 in $gi$ filters
were affected by $3^{\prime\prime}$ seeing, pointing 3 by
$2^{\prime\prime}$. This resulted in a 
degradation of image depth by 0.8 and 1.0 mag for $g$ and $i$,
respectively, as compared to the values listed in Table
\ref{obstable}. The data are still deep enough to probe the
the red sequence at $z=0.45$. However, the very large seeing
differences between filters $gri$ made accurate colour measurements
for pointings 1 and 2 difficult despite our attempt to homogenise the
PSF, resulting in reduced number densities for these pointings.

The astrometric and relative photometric calibration was performed
similar to the CFHT images, calibrating all images
simultaneously. Since the illumination correction for WFC/INT is 
unknown, we determined the ZPs for each of the four detectors 
independently through direct comparison with SDSS. Variations between 
chips are $\leq0.02$ mag. Only the southernmost $10^\prime-20^\prime$
of pointings $3-6$ are not covered by SDSS (Fig. \ref{fields}).
Detectors in these areas were calibrated using average correction
terms determined from the other pointings with full SDSS overlap. In
this way homogeneous photometry is achieved across all 12
pointings. For better comparison with the CFHT data we determined
average corrections from the overlap areas,
\begin{equation}
\langle g_{\rm CFHT} - g_{\rm INT}\rangle = -0.09\;{\rm mag}
\end{equation}
\begin{equation}
\langle r_{\rm CFHT} - r_{\rm INT}\rangle = +0.07\;{\rm mag}
\end{equation}
\begin{equation}
\langle i_{\rm CFHT} - i_{\rm INT}\rangle = +0.14\;{\rm mag}
\end{equation}
and applied these to the INT zeropoints.

As can be seen from Fig. \ref{fields}, the sky coverage with WFC/INT
is not complete due to the asymmetric detector layout. We checked the
corresponding areas in SDSS to ensure that we did not
miss any potential further member clusters.

\begin{table}
\caption{Characteristics of the MPG-ESO/WFI, CFHT/Megaprime and 
INT/WFC data. The limiting AB magnitudes represent the $5\sigma$
completeness limit for non-stellar sources. The scattering in seeing
and depth for the INT/WFC pointings is given as well (excluding
pointings $1-3$ for $gi$ which are significantly worse, see
Sect. \ref{intimaging}).}
\label{obstable}
\begin{tabular}{l r r r r}
\hline
\hline
\noalign{\smallskip}
Instrument & Filter & $t_{\rm exp}$ [s] & Seeing [\myarcsecnodot] &
$M_{\rm lim}$ [AB]\\
\hline
\noalign{\smallskip}
ESO/WFI & $R$ & 7200 & 0\myarcsec83 & 24.7 \\
\hline
\noalign{\smallskip}
CFHT/Megaprime & $u$ & 4250 & 0\myarcsec99 & 25.0\\
CFHT/Megaprime & $g$ & 3000 & 0\myarcsec76 & 25.4\\
CFHT/Megaprime & $r$ & 5000 & 0\myarcsec66 & 25.4\\
CFHT/Megaprime & $i$ & 6000 & 0\myarcsec57 & 25.2\\
CFHT/Megaprime & $z$ & 5000 & 0\myarcsec73 & 23.3\\
\hline
\noalign{\smallskip}
WFC/INT & $g$ & 1800 & 1\myarcsec50$\pm0.21$ & 24.2$\pm0.2$ \\
WFC/INT & $r$ & 1800 & 1\myarcsec33$\pm0.14$ & 23.5$\pm0.1$ \\
WFC/INT & $i$ & 1800 & 1\myarcsec30$\pm0.13$ & 22.8$\pm0.1$ \\
\hline
\end{tabular}
\end{table}

\subsection{Catalogues}
\subsubsection{Photometric catalogues and masking}
The photometric catalogues were created as follows. After PSF
homogenisation we used SExtractor \citep{bea96} in double image mode,
using the $i$-band image for the detection channel and the individual
filter stacks for photometry (for the INT data a noise-normalised
stack of the $gri$ images served as the detection channel). In this
way the object flux in each filter was integrated over identical
apertures. We kept objects with at least 5 connected pixels with
$S/N\geq 1.5$ each. Objects near brighter stars were removed using the
{\tt automask} tool \citep{del07,ehl09}.

\subsubsection{Shear catalogue}
The shear catalogue is based on a noise-normalised combination of
the CFHT $r$- and $i$-band images for extra depth. SExtractor
detections with a minimum number of 5 connected pixels with 
${\rm S/N}\geq2.0$ form the primary catalogue. The latter was
fed into our implementation \citep[][]{ewb01} of the KSB method
\citep{ksb95,luk97,hfk98} for shape measurement. A description of the
PSF correction can be found in \cite{bas01}.

Objects with a low KSB detection significance ($\nu_{\rm max}<10$) or
a PSF corrected modulus of the ellipticity larger than 1.5 are removed
from the catalogue (the ellipticity can become larger than 1 due to
the PSF correction). We also reject galaxies for which the correction
factor $({\rm Tr}\,P^{\rm g})^{-1}>5$ \citep[][]{ewb01}. More details
about this can be found in \cite{seh07}.

To remove unlensed foreground objects we kept galaxies with
$z_{\rm phot}>0.5$ only. The number density of lensed galaxies
after filtering is $n=19.5$ arcmin$^{-2}$, and the width of the 
ellipticity distribution $\sigma_{|\varepsilon|}=0.37$. Both $n$
and $\sigma_{|\varepsilon|}$ 
determine the $S/N$ of the weak lensing analysis. The
unweighted mean (median) photometric redshift of the shear catalog is
$\langle z\rangle=1.15$ (1.00).

\subsection{\label{emmi}NTT/EMMI spectroscopy}
Multi-object spectroscopy for the three most promising shear-selected 
clusters in the MPG-ESO/WFI data was taken with NTT/EMMI (ESO/La
Silla) in 2007-08-21/22 in sub-arcsecond seeing conditions. We covered
the $3800\text{\AA}-9500\text{\AA}$ wavelength range with a spectral
resolution of $3\myarcsec6\,\text{\AA}\,{\rm pixel^{-1}}$. About 40
slitlets, 1\myarcsec3 wide, were placed in each of the three slit
masks. Exposure times were $6\times1800$s for the northern-most field,
and $4\times1800$s for the other two pointings.

The data were reduced using a custom pipeline which
will be described in a forthcoming paper. In short, the data were
overscan corrected, debiased and flat-fielded. We corrected for
spatial distortions of the spectral footprints and small rotations of
the slitlets, and subtracted a background model. Wavelength
calibration was done based on combined Helium/Argon arc lamps. The
pre-processed spectra were combined using a weighted average, and
1D-spectra were extracted. Flux-calibration and correction for
telluric absorption were not performed. The redshift determination was
done using [OII] 3728, H$\beta$ 4862, [OIII] 5007 and H$\alpha$ 6564
emission lines, as well as CaII 3934/3969, $G$-band (4305\AA), Mg 5177
and $E$-band (5269\AA) absorption features. 

\section{Measuring mass and light in SCL2243}
\subsection{\label{tracer}Selection and distribution of early-type 
  galaxies}
Early-type galaxies form preferentially in high-density cluster 
environments through interaction with neighbours and the stripping
of intra-galactic gas \citep[see e.g.][]{qmb00,pef09,sws09}. However,
observations show that star formation is already
enhanced in much less dense environments at clustercentric distances
of several virial radii, consuming the gas available for further star
formation. The latter is then subdued quickly with
increasing proximity to denser clusters and filaments. Therefore we 
expect to find an increasing fraction of passive redder galaxies with 
colours similar to those of early-type galaxies in regions with
enhanced mass density
\citep[see][for further details and observational evidence]
{gzb04,por07,thk07,prp08,vzg08}.
Consequently, by selecting galaxies with early-type colours we should be
able to not only identify the main clusters in SCL2243, but
also to map the connecting skeleton over large distances. This turned
out to work well as we show in the following. In a second step, we
create a 2-dimensional map of the luminosity density
(Sect. \ref{lummap}) based on the distribution and fluxes of the
sample of early-type galaxies, and use it as a guide for our weak
lensing mass integration along the filaments (see
Sect. \ref{filamentlensing} for details).

\begin{figure}[t]
  \includegraphics[width=1.0\hsize]{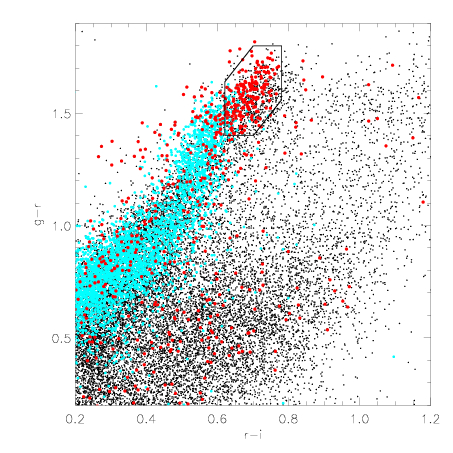}
  \caption{\label{scl2243_gry_colmag}{$g-r$ vs. $r-i$ colour-colour
      diagram. Red dots: Galaxies with distances smaller than
      $1.8$ Mpc from of the supercluster centre. The box outlines
      our photometric selection criteria for early-type galaxies
      (equations \ref{selcrit1} to \ref{selcrit3}) in the
      supercluster. Blue dots: Galaxies with 
      $0.27\leq z_{\rm phot}\leq 0.39$ from the full CFHT field.}}
\end{figure}

Our sample of elliptical galaxies is extracted from a $g-r$
vs. $r-i$ colour-colour diagram. We overplot member galaxies of 
MACS J2243 (Fig. \ref{scl2243_gry_colmag}, red dots) and in this way
determine the selection criteria for early-type galaxies as
\begin{equation}
\label{selcrit0}
0.40 \leq z_{\rm{phot}} \leq  0.52
\end{equation}
\begin{equation}
\label{selcrit1}
1.40 \leq g-r \leq 1.80
\end{equation}
\begin{equation}
\label{selcrit2}
0.62 \leq r-i \leq 0.78
\end{equation}
\begin{equation}
\label{selcrit3}
2.0\,(r-i) \leq g-r \leq 2.0\,(r-i) + 0.4\,.
\end{equation}
Equation (\ref{selcrit0}) covers the $1\sigma$ uncertainty around the
mean photometric cluster redshift of $0.46$ (slightly higher than the
spectroscopic value of 0.45). The remaining criteria are visualised in
Fig. \ref{scl2243_gry_colmag}. In addition, we adopted a lower limit
of $i\leq 22$ mag ($M_i\leq -19.3$ at $z=0.45$) to reject high-z
interlopers. Note that these criteria are partially redundant as
equations (\ref{selcrit1}) to (\ref{selcrit3}) already perform a good
selection in redshift, sufficient to select mainly early-type galaxies
in the desired redshift range for the 12 INT/WFC pointings. The latter
were observed in $gri$ filters only and thus have no photometric
redshifts. 2507 early-type galaxies with $i\leq 22$ and suitable
photometric redshifts (respectively colours) remain in the full field, 
857 of which lie inside the CFHT field. For comparison, inside the
CFHT area 1664 late-type galaxies with $0.40 \leq z_{\rm{phot}} \leq
0.52$ and $i\leq 22$ are found.

\begin{figure*}[t]
  \includegraphics[width=1.0\hsize]{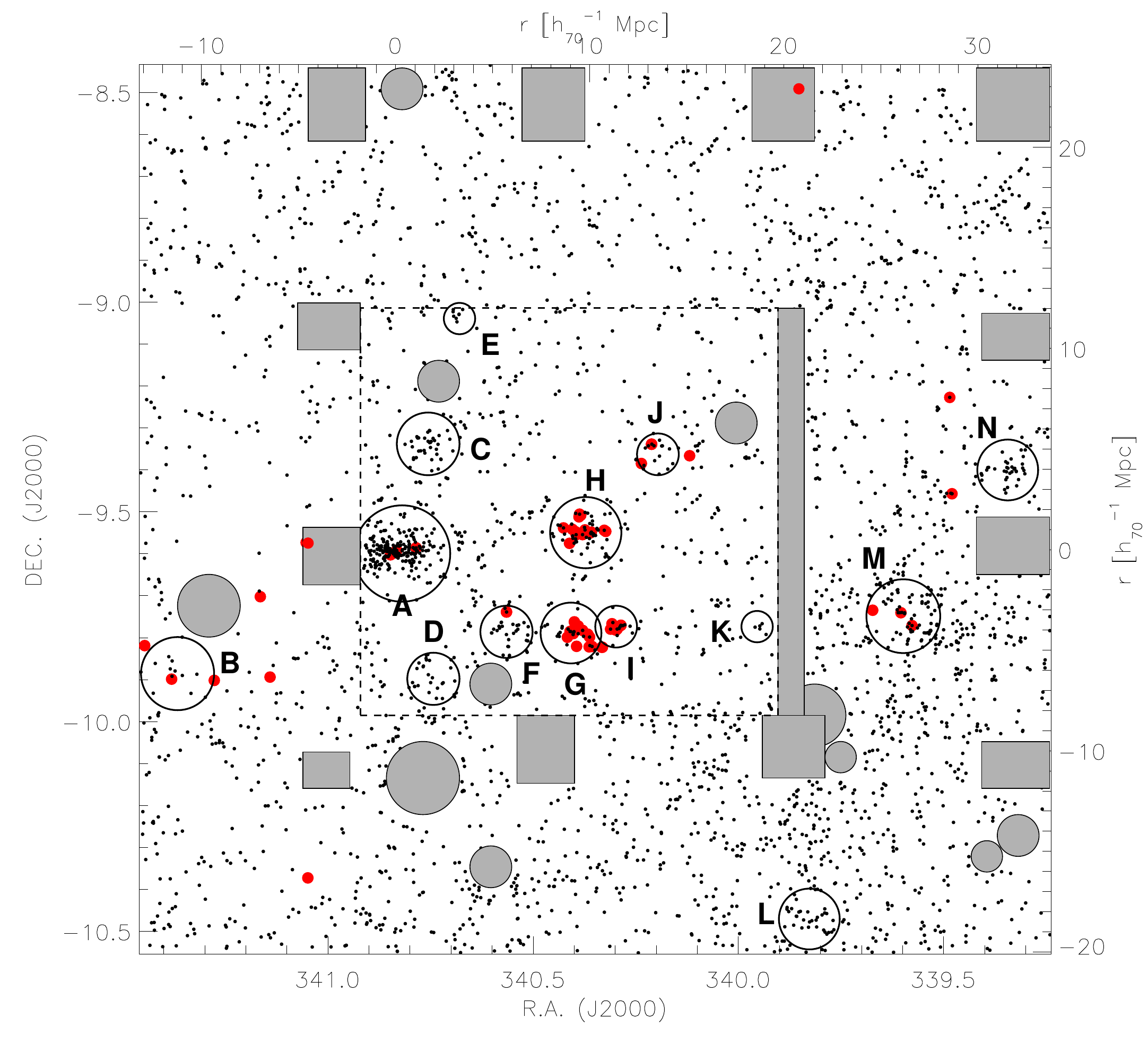}
  \caption{\label{scl2243_fullfield}{Distribution of early-type
      galaxies with $z\sim0.45$. Sky coverage is incomplete, as is
      indicated by the shaded areas (rectangular: gaps due to
      telescope pointing, circular: bright stars). The dashed line
      indicates the CFHT field of view. The reduced number density of
      objects left of the CFHT field is caused by strongly varying
      seeing conditions which made precise colour measurements very
      difficult (see Sect. \ref{intimaging} for details). Empty
      circles mark member clusters of overdensities, their size
      qualitatively indicating richness.}}
\end{figure*}

The spatial distribution of these galaxies is shown in Fig. 
\ref{scl2243_fullfield} (small black dots). Large red dots indicate
galaxies for which we have matching SDSS spectra (full field) and
EMMI spectra (peaks G, H, I, and J). The 
central cluster (MACS J2243, labelled \textit{A}) forms a highly
significant overdensity. Within the CFHT field we select 10 clusters
or overdensities, based on a minimum peak luminosity density of 
$10^{10.1}\,L_\odot\,{\rm Mpc}^{-2}$. For WFC/INT we present 
the four most promising clusters only, as part of the area
is contaminated with foreground structures at $z=0.36-0.39$. We cannot
unambiguously separate these from smaller overdensities at 
$z\sim0.45$. Our cluster detections are listed in Table
\ref{clustertable}. Half of these do not form typical clusters but 
loose associtations of galaxies (see also the images in the
Appendix).

\subsection{\label{n200}Luminosity-based $r_{200}$ and $M_{200}$ of
  MACS J2243}
K-corrections and rest-frame luminosities in $i$-band were calculated
using {\tt kcorrect} v4.2 \citep{blr07}. As we have projected the
filter transmission curves to the main cluster redshift of $z=0.45$,
we denote them as e.g. $^{0.45}i$. Since {\tt kcorrect} evaluates
luminosities for $h_{100}=1.0$, we corrected the absolute magnitudes by 
$5 {\rm log}_{10}(h_{100}=0.7)\sim-0.77$ mag.

\cite{hmw05} and \cite{jsw07} have shown that $r_{200}$ and $M_{200}$
can be estimated from the number $N_{\rm gal}$ of galaxies
within a radius of $1\,h_{100}^{-1}$ Mpc of the brightest cluster
galaxy (BCG). Only galaxies in the red sequence and with $i$-band
luminosities $L>0.4 L_*$ are considered. The refined version from the
relation of \cite{hmw05} reads 
\begin{equation}
\label{hansen05}
r_{200}^{\rm gal}=0.156\, h_{100}^{-1}\; {\rm Mpc}\; N_{\rm gal}^{0.60}\,,
\end{equation}
being the radius within which the luminosity is 200 times the mean
luminosity of the Universe. It must not be mistaken for $r_{200}$
which refers to matter overdensity. Based on weak lensing measurements
and the number $N_{200}$ of galaxies within $r_{200}^{\rm gal}$,
\cite{jsw07} and \cite{hsw09} obtain
\begin{equation}
r_{200}=0.182\,h_{100}^{-1}\,{\rm Mpc}\,N_{200}^{0.42}
\label{sr_r200}
\end{equation}
\begin{equation}
M_{200}=1.75\times10^{12}\,h_{100}^{-1}\,{\rm M_\odot}\,N_{200}^{1.25}\,.
\label{sr_m200}
\end{equation}

\begin{table}
\caption{List of possible member clusters and sub-structures. $r$ is
  the angular separation from the supercluster centre, 
  $\langle z\rangle$ the mean redshift, and $N(z)$ the number of
  spectroscopically confirmed member galaxies.}
\label{clustertable}
\begin{tabular}{l r r r r r}
\hline
\hline
Cluster ID & RA (J2000) & DEC (J2000) & r [$^{\prime}$] & $\langle z\rangle$ & N(z)\\
\hline
SCL2243-A & 22:43:20 & -09:35:25 & 0.0 & 0.447 & 36$^a$\\
SCL2243-B & 22:45:32 & -09:53:55 & 37.4 & 0.440 & 1\\
SCL2243-C & 22:43:01 & -09:19:30 & 16.6 & $-$ & $-$\\
SCL2243-D & 22:42:54 & -09:52:20 & 18.7 & $-$ & $-$\\
SCL2243-E & 22:42:45 & -09:02:10 & 34.3 & 0.408 & 1\\
SCL2243-F & 22:42:16 & -09:44:50 & 18.6 & 0.437 & 1\\
SCL2243-G & 22:41:38 & -09:47:20 & 28.6 & 0.436 & 14\\
SCL2243-H & 22:41:27 & -09:33:00 & 28.6 & 0.447$^{b}$ & 15\\
SCL2243-I & 22:41:09 & -09:46:20 & 34.6 & 0.436 & 6\\
SCL2243-J & 22:40:42 & -09:22:10 & 41.7 & 0.456$^{c}$ & 3\\
SCL2243-K & 22:39:53 & -09:48:05 & 53.5 & $-$ & $-$\\
SCL2243-L & 22:39:06 & -10:29:15 & 82.5 & $-$ & $-$\\
SCL2243-M & 22:38:17 & -09:45:00 & 75.2 & 0.432 & 2\\
SCL2243-N & 22:37:21 & -09:23:55 & 89.1 & $-$ & $-$\\
\hline
\end{tabular}
$^{a}$: MACS J2243-0935 \citep{eem10}\\
$^{b}$: Superimposed group at $z=0.256$\\
$^{c}$: Superimposed cluster at $z=0.326$\\
\end{table}

\begin{figure}[t]
  \includegraphics[width=1.0\hsize]{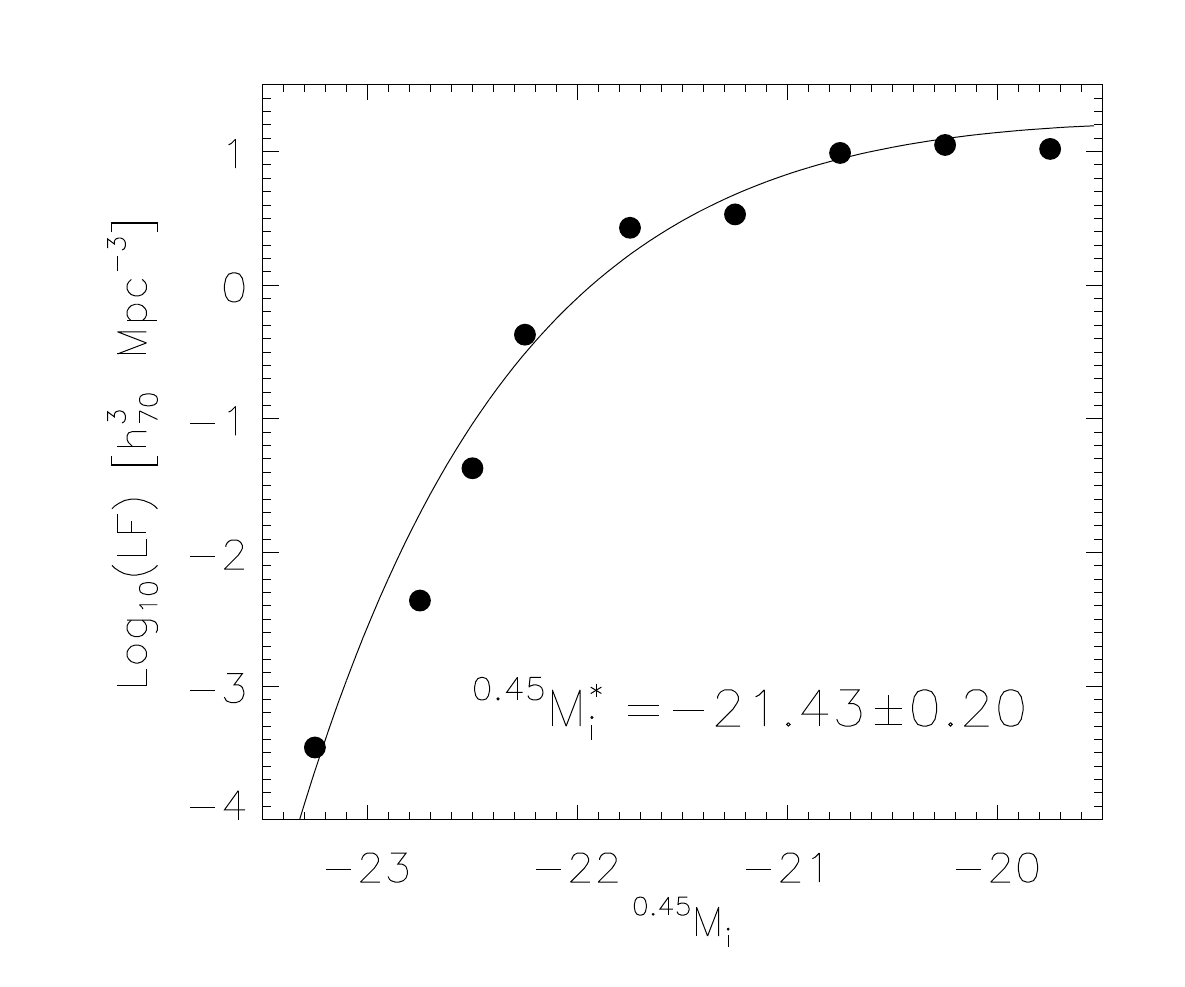}
  \caption{\label{lumfunc}{Rest-frame $^{0.45}i$-band luminosity
      function for the early-type galaxies within $r_{200}=2.06$ Mpc
      of the supercluster centre. The solid line shows the best-fit
      Schechter model.}}
\end{figure}

Fitting a Schechter-type luminosity function \citep{sch76} to the
early-type galaxies in MACS J2243, we determine 
$M_{^{0.45}i}^*=-21.43\pm0.20$ mag (see Fig. \ref{lumfunc}) and
$N_{200}=150^{+31}_{-29}$. Using equations (\ref{sr_r200}) and
(\ref{sr_m200}) we find for the central cluster MACS J2243
\begin{equation}
r_{200} = 2.13^{+0.18}_{-0.12}\;h_{70}^{-1}\;{\rm Mpc}\,
\end{equation}
\begin{equation}
M_{200} = 1.31^{+0.25}_{-0.20}\times 10^{15}\;h_{70}^{-1}\,M_\odot\,
\end{equation}
\begin{equation}
L_{200} = 3.60^{+0.10}_{-0.12}\times 10^{12}\;h_{70}^{-2}\,L_{i,\odot}\;.
\end{equation}

\subsection{\label{lummap}The luminosity map of the central square degree}
For our analysis of the light distribution in SCL2243 we smooth
the luminosity field of the early-type galaxies with a 
${\rm FWHM}=4^{\prime}$ wide Gaussian kernel, corresponding to a
physical scale of $1.38\,h_{70}^{-1}$ Mpc at $z=0.45$. The kernel is
truncated at a radius of $2\sigma$ and has a constant value subtracted
such that it becomes equal to zero at $r=2\sigma$ (and it is
normalised). In this way we suppress the effects of the broad
wings. This kernel is significantly smaller than the smoothing radius of 
$8h_{100}^{-1}$ Mpc chosen by \cite{eet07} in their search
for superclusters (using an Epanechnikov kernel which has steeper edges
than our Gaussian filter). The purpose of their work was to identify
superclusters in a large volume, whereas we are interested in the
internal structure of SCL2243.

We show the luminosity density of the CFHT field in Fig. 
\ref{scl2241_redsequ_specz_045}, overlaid as green contours over the
distribution of early-type galaxies (black dots). Contour levels
increase in logarithmic steps of 0.3. Red dots are spectroscopically
confirmed member galaxies, three from SDSS in the main cluster, and
all but two of the rest from NTT/EMMI. Dots with cyan colours belong
to a foreground cluster at $z=0.326$. Blue contours represent the weak
lensing mass reconstruction (see Sect. \ref{massrec}).

SCL2243-J (see Table \ref{clustertable}) is masked from
all subsequent analyses at it is projected behind a 
foreground cluster at $z=0.326$. We recognise three filaments
connected to MACS J2243, which we call AC, AH and AFDGI
according to the labelling we chose for the sub-structures
identified (Fig. \ref{scl2243_fullfield}). These structures extend
over $(5-15)\,h^{-1}$ Mpc. We predict at
least one more filament South-East of MACS J2243, connecting to
SCL2243-B (Fig. \ref{scl2243_fullfield}). The characteristic
width of all filaments is about $2.0\,h_{70}^{-1}$ Mpc, with local 
enhancements up to $\sim3.0$ Mpc where concentrations of galaxies or
small member clusters are found.

We measure a total luminosity of early-type galaxies in the CFHT field
of $L^{\rm tot}_{^{0.45}i}=1.43\times10^{13}\,L_\odot$. Within $r_{200}$
the luminosity of MACS J2243 is $(3.60\pm0.11)\times10^{12}L_\odot$,
whereas filaments AC, AH and AFDGI contribute 
$(5.19\pm0.31)\times10^{12}L_\odot$.

\subsection{Colour variations within the supercluster structure}
Figure \ref{colormap} displays the unweighted $\langle g-i\rangle$
colours of all galaxies with $0.40\leq z_{\rm phot}\leq 0.52$ and $i<22$
mag. For visual reference we overlaid the luminosity contours of the
early-type galaxies tracing the supercluster structure. MACS J2243 is
the most massive object and thus also the one with the reddest
colour ($g-i=2.27$ mag) at its centre. At a clustercentric
distance of $0.25\times r_{200}$, $g-i$ starts decreasing, reaching
$1.90\pm0.05$ mag at the virial radius.

This is not much redder than filaments AC and AFDGI with
$\langle g-i\rangle=1.80$ (0.05 mag scattering). These filaments
become redder by about 0.05 mag in areas of greater density (objects
C, D, F, G, and I), but are otherwise remarkably constant out to
distances as large as $6\times r_{200}$ (even $8\times r_{200}$ or 18
Mpc if SCL2243-K is spectroscopically confirmed). Also, the colour of
different filaments is the same with one exception: the middle of
filament AH is noticably bluer, $\langle g-i\rangle=1.71\pm0.02$ mag.  

Filaments AH and AF merge before connecting to the western-most side
of cluster A. This area is significantly redder by 0.08 mag forming a
cluster infall region. Objects E, H and J appear further evolved than
the rest of the field with $g-i=2.0$ mag (E, H) and 1.9 mag (J),
respectively. 

\subsection{\label{massrec} The mass map of the central square degree}
Weak gravitational lensing has been used with great success when mapping
the matter distribution and density profiles of clusters of galaxies
\citep[see][for a listing]{dah07}. In this paper we use the model-free
finite-field method from \cite{ses01} to reconstruct the dimensionless
surface mass density, $\kappa$. This method uses the field border as a
boundary condition and is easily implemented for rectangular data
fields\footnote{Our implementation of this algorithm is available at\\
http://www.astro.uni-bonn.de/$\sim$mischa/download/massrec.tar .}.

Mass reconstruction algorithms must operate on smoothed shear
fields as otherwise noise contributions become infinite
\citep{kas93}. We chose the same truncated Gaussian filter
as for the luminosity field (Sect. \ref{lummap}), with an
identical width of 4\myarcmin0 and a constant value subtracted
such that the filter becomes zero at the maximum radius of
$2\sigma$. This zero-setting of the filter is a requirement for the
\cite{ses01} algorithm as it ensures that the derivatives of the
smoothed shear field are well-behaved.

\begin{figure*}[t]
  \includegraphics[width=1.0\hsize]{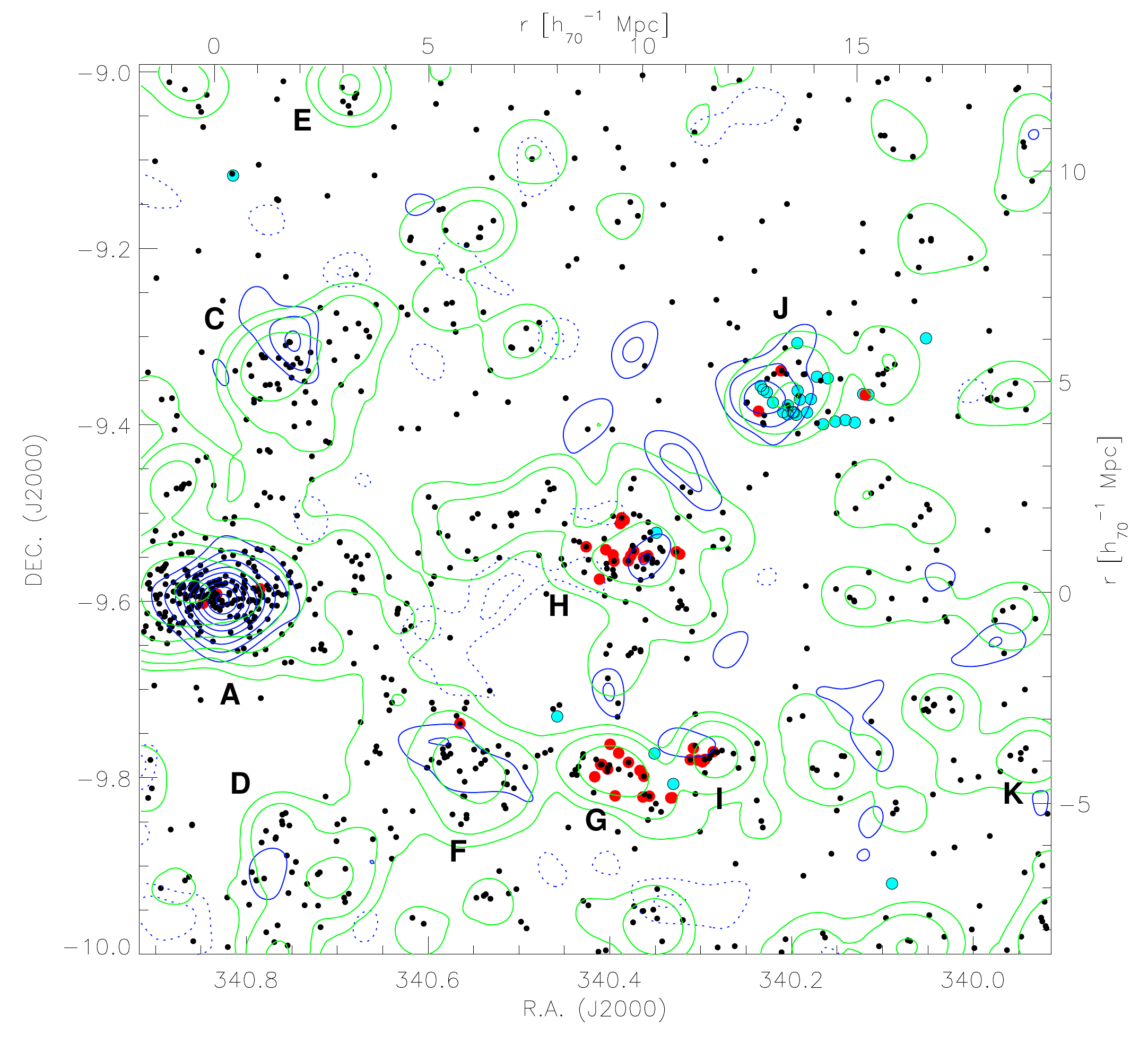}
  \caption{\label{scl2241_redsequ_specz_045}{
      Early-type galaxies with $0.40\leq z_{\rm{phot}}\leq 0.52$ (black
      dots). Overlaid are galaxies with EMMI and SDSS spectra within
      $0.43<z_{\rm{spec}}<0.46$ (red dots). Blue dots with black
      circles have $0.32<z_{\rm{spec}}<0.33$. The cluster at
      RA$=340.2$ DEC$=-9.35$ is a foreground object ($z=0.326$) 
      superimposed over a sub-structure at $z=0.456$. The solid blue
      contours represent the S/N of the mass reconstruction, starting
      at $2\sigma$ and increasing in steps of $1.0\sigma$; dashed
      contours represent negative values ($2\sigma$ and $3\sigma$). 
      Green contours trace the smoothed luminosity density $I$ (in 
      $L_{i,\odot}\,{\rm Mpc}^{-2}$) of the early-type member galaxies
      in logarithmic units, starting with Log$_{10}(I)=9.6$ and
      increasing in steps of 0.3. The smoothing length for the surface
      mass density and luminosity density was $4\myarcmin0$. The upper 
      and right axes give the physical scale at the main cluster
      redshift of $z=0.447$. Compare to Fig. \ref{scl2241_blue} in the
      Appendix for late-type galaxies in the same redshift range.}}
\end{figure*}

The reconstruction algorithm only works for under-critical
regions with $\kappa<1$, i.e. strong lensing areas are not
reconstructed reliably. This is of no concern for our work as the
smoothing scale is significantly larger than the strong lensing
regions in the core of MACS J2243.

The convergence $\kappa$ is determined up to an additive constant, the
so-called `mass-sheet degeneracy', which is broken by assuming that
$\kappa$ vanishes on average along the border of the field. We
excluded part of the edge close to MACS J2243 to avoid biasing by
increased values of $\kappa$ in this area. The
uncertainty in the determination of the mass sheet degeneracy can be
estimated from the variations along the edge in randomisations
(see below) and the number of independent apertures that can be placed
along the border. The uncertainty of the mass sheet degeneracy
correction term is $\sigma_\kappa=0.0023$.

\begin{figure*}[t]
  \includegraphics[width=1.0\hsize]{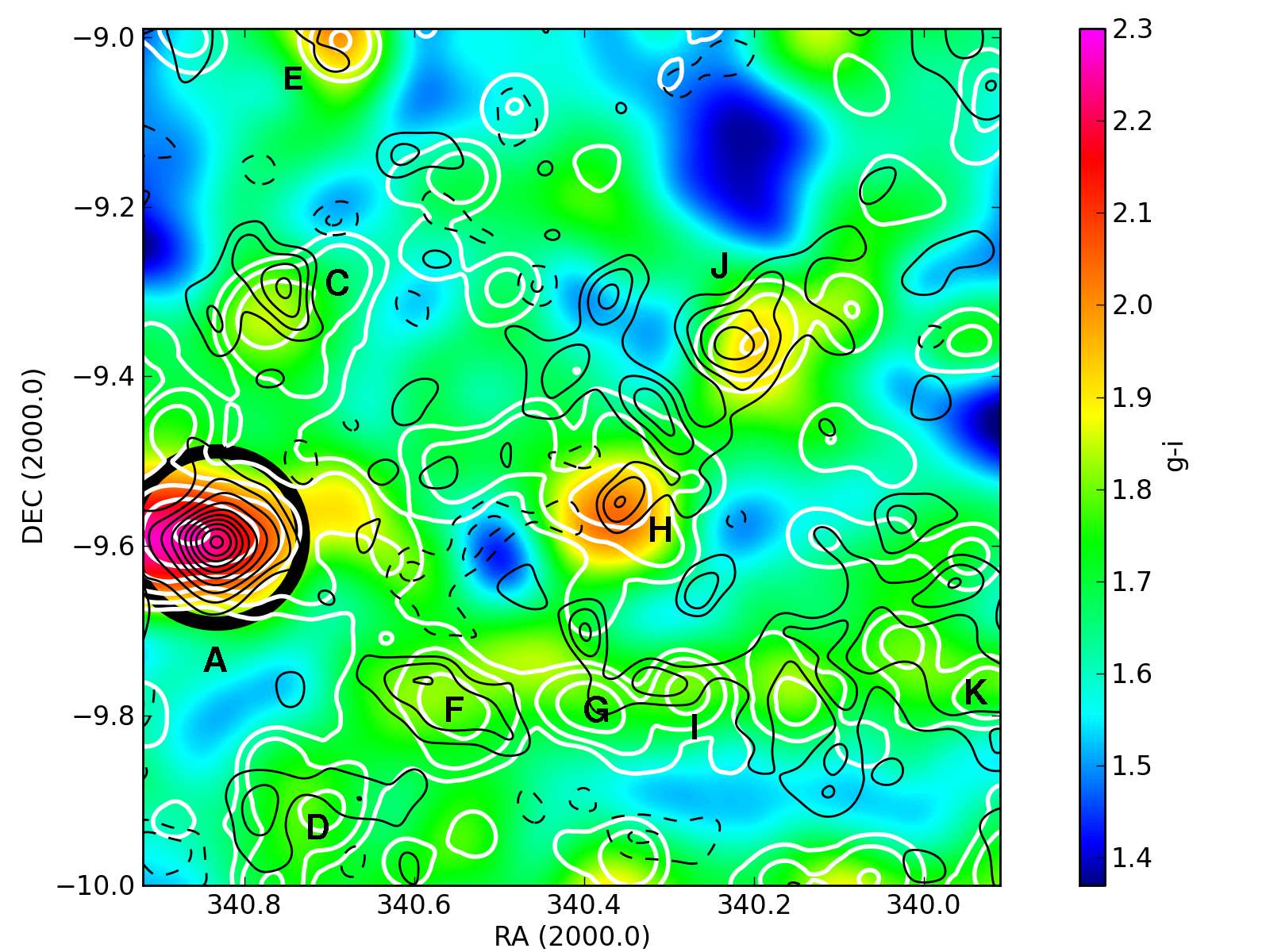}
  \caption{\label{colormap}{The $g-i$ colours of all
      galaxies (early- and late-type) with $0.40\leq z_{\rm phot}\leq0.52$.
      For visual reference we overlaid the luminosity contours of 
      the early-type galaxies tracing the structure of the
      supercluster, and the mass reconstruction (thin solid black 
      lines, this time starting at $1\sigma$). The black dashed
      contours represent negative mass reconstruction levels
      (2$\sigma$ and 3$\sigma$). The thick black circle marks
      $r_{200}$, its width indicates the $1\sigma$ uncertainty.}}
\end{figure*}

The noise level of the mass reconstruction mainly depends on two
quantities: the width of the distribution of intrinsic ellipticities
($\sigma_{|\varepsilon|}=0.37$) and the local number density of
background galaxies. The latter would in principle be fairly constant
if it was not significantly altered by the masks
we put on top of brighter stars. To obtain the noise map, we created
1000 realisations of randomised galaxy orientations keeping their
positions fixed and reconstructed the $\kappa$-field for each of them.
The two-dimensional rms of these $\kappa$-maps then yields the desired
noise map. Since lensing increases the ellipticities of galaxies, we
removed the radially averaged tangential shear profile of MACS J2243
prior to the randomisations. Otherwise the noise at the cluster
position would be overestimated. This correction is only significant
for the supercluster core (improving its S/N by about 10\%), and
negligible for all other mass concentrations as their noise is
dominated by intrinsic ellipticities. The resulting mass map is
overplotted (blue respectively black contours) over the luminosity and 
colour maps shown in Figs. \ref{scl2241_redsequ_specz_045} and
\ref{colormap}.

We detect MACS J2243 at the $10\sigma$ level and estimate
\begin{equation}
r_{200} = (2.06\pm0.13)\;h_{70}^{-1}\;{\rm Mpc}
\end{equation}
\begin{equation}
M_{200} = (1.54\pm0.29)\times 10^{15}\;M_\odot\;.
\end{equation}
This is in very good agreement with our earlier luminosity-based
analysis in Sect. \ref{n200}, yielding
$r_{200} = 2.13^{+0.18}_{-0.12}\;h_{70}^{-1}\;{\rm Mpc}$ and 
$M_{200} = 1.31^{+0.25}_{-0.20}\times 10^{15}\;h_{70}^{-1}\,M_\odot$,
respectively.

\begin{figure*}[t]
  \includegraphics[width=1.0\hsize]{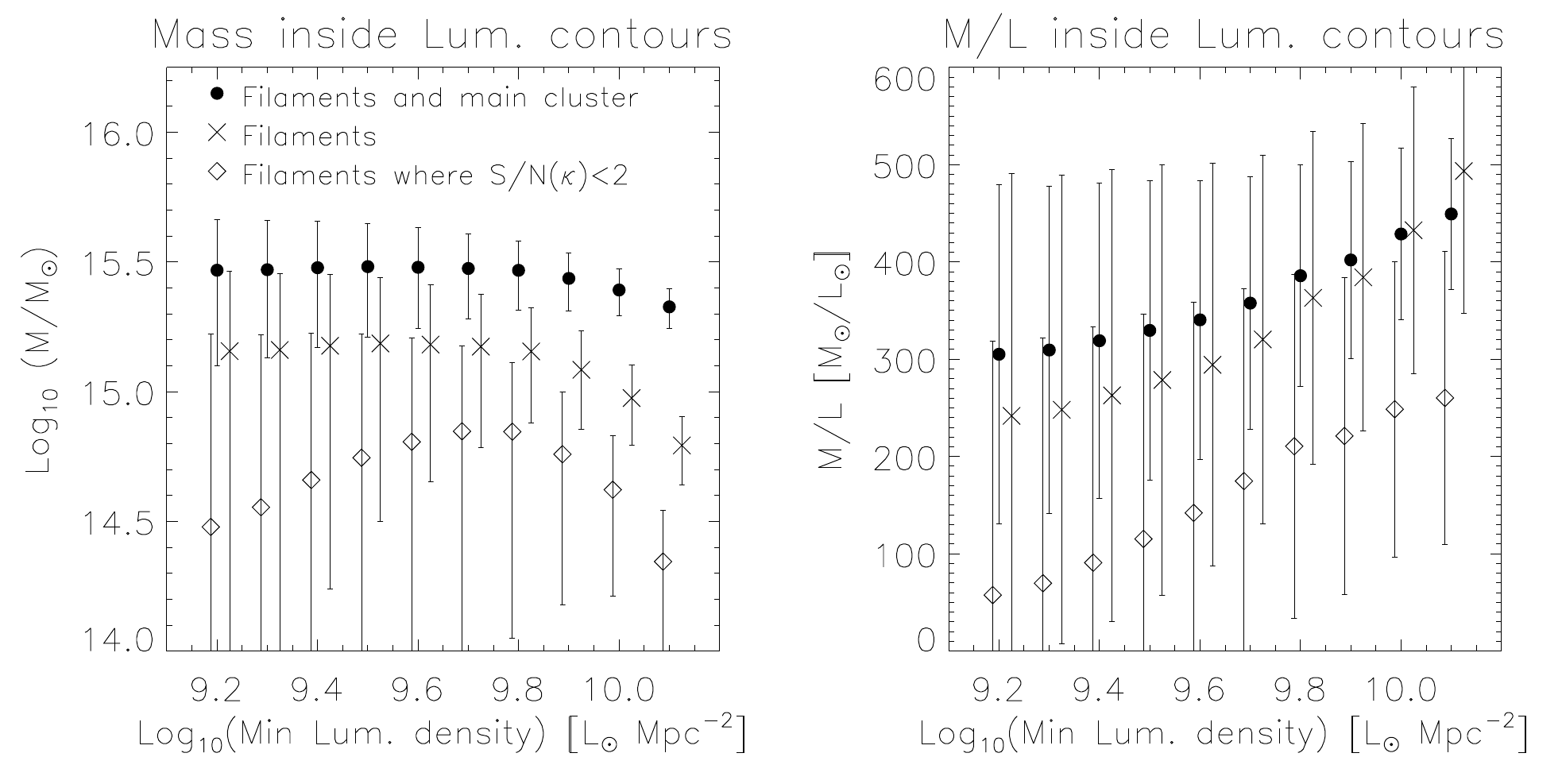}
  \caption{\label{scl2241_ml}{Filaments have not yet been detected
      directly with weak lensing. However, we can attempt to integrate
      the weak lensing reconstructed surface mass density $\kappa$
      within different isophotes, yielding a statistically significant
      measurement due to the increased area. Left: The total
      mass in the filaments obtained by integrating $\kappa$, with and
      without condensations in the filaments (crosses and diamonds,
      respectively). For comparison we show the total,
      including the supercluster centre (black dots). More details in
      the text. Error bars were calculated from the 1000 randomised
      data fields. Right: Same as the left panel, but for the M/L
      ratio.}}
\end{figure*}

\subsection{\label{filamentlensing}Weak lensing and filaments}
Weak gravitational lensing has so far been hitting a limit when it
comes to the direct detection of filaments, as their density is 
significantly lower and their extent much larger than those of 
galaxy clusters. \cite{gtm02} reported a ground-based  
weak lensing detection of a small filament in A901/902, which could
not be confirmed by HST/ACS observations \citep{hgp08}.
Likewise, \cite{kwl98} found a filament in MS0302, but argued it could
be due to projection effects from structures elsewhere along the line
of sight. This feature was not seen by \cite{gmf04}. Similarly,
\cite{dsc05} reported a weakly significant lensing detection of a
filament connecting the close double clusters A222/223. They have
argued, however, that it is difficult to disentangle the shear signals
of the close pair of clusters from that of the filament. 

Note that all these filaments are very small with lengths on the order
of $1\,h_{70}^{-1}$ Mpc, connecting very close clusters, whereas the
filaments we study in SCL2243 are approximately one order of magnitude
larger, representing the actual cosmic web.

\cite{mkm10} have discuss the subject of (cosmic web) filament weak
lensing based on simulations. They have investigated three possible
measurement methods, one of them being the direct approach of
model-free mass reconstruction using the same technique as we do in
this paper. They have found that filaments are in general difficult to
detect given their low surface mass density ($\kappa\sim 0.01$). With
number densities of 30 galaxies arcmin$^{-2}$ typical for very deep
ground-based imaging, even the most massive filaments were only weakly
detected. Space-based observations yield about three times higher
number densities, which improve detection statistics.

\subsubsection{Weighing supercluster filaments with weak lensing} 
The direct detection of filaments in a supercluster with weak lensing
is a difficult task and has so far not been achieved. This also holds
for our data set, as we only detect the most massive concentrations in
the filaments with $S/N\sim3-4$ in the mass reconstruction.

However, absence of detection does not mean absence of signal. As we
have summarised at the beginning of Sect. \ref{tracer}, early-type
galaxies serve as a good tracer for the underlying distribution of
gravitating matter. Therefore one would expect a trend towards
positive values of $\kappa$ wherever early-type galaxies dominate the
field. Indeed we see a correlation of $\kappa$ with the
galaxy distribution at the $S/N\sim1-2$ contour level, in particular 
along the filament DFGIK, and in general where the galaxy population
has colours larger than about 1.7 (Fig. \ref{colormap}).

At a given point on sky this is of course insignificant. But by
integrating $\kappa$ over areas where the smoothed luminosity or 
colour exceeds a certain value, we can arrive at a statistically
significant measurement. This is equivalent to the statistical
stacking of weak lensing signals, in order to make a measurement that
would otherwise be unfeasible, or to obtain high-S/N statistics
\citep[e.g.][]{hfk01,phc05,sjf04,sjm09}. 
%Using this approach we reach
%an effective number density of $n\sim180$ galaxies arcmin$^{-2}$ as
%compared to the actual value of $\sim19$ arcmin$^{-2}$. This is a very
%crude estimate for illustration purposes only, obtained from the
%approximate number of independent apertures which we can place on the
%filaments.
The results of this measurement are shown in Fig. \ref{scl2241_ml},
where we integrated $\kappa$ over areas with different
isophotes. Our analysis is restricted to the region covered 
by the supercluster centre and filaments AC, AH, and AFDGI. We have
repeated it three times, integrating over
\begin{itemize}
\item{the entire supercluster}
\item{the filaments only}
\item{the filaments only, excluding mass concentrations with 
  $S/N(\kappa)>=2$.}
\end{itemize}
To better understand the plots in Fig. \ref{scl2241_ml} one must note
that for very low isophotal boundaries one integrates over a larger
area. This increases the noise and also includes underdense regions
with negative $\kappa$. This is felt mostly by the naked filaments 
(i.e. without contribution from embedded clusters or overdensities),
thus the integrated mass for the lower curve (left panel, diamonds)
decreases. Isophotes brighter than 
$\sim 10^{9.8-9.9} L_\odot\,{\rm Mpc}^{-2}$ cover increasingly smaller
areas and thus smoothing effects become important, distributing mass
and light to areas outside those contours. Of course the smaller
integration area also leads to decreased mass estimates. In addition,
due to characteristic positional uncertainties of 
$1^{\prime}-2^{\prime}$ in the mass reconstruction 
\citep[due to noise, see e.g.][]{seh07}, the surface mass density
field starts to decouple from the luminosity field. As a consequence
of the latter three effects, mass estimates decrease significantly if
the integration contour is too tight (higher than about 
$\sim 10^{9.8-9.9} L_\odot\,{\rm Mpc}^{-2}$ for our data). The same effects
hold for the $M/L$ ratio (right panel of Fig. \ref{scl2241_ml}).
For the discussion below (Sect. \ref{filamentmass}) we constrain our 
analysis of the filaments' mass and $M/L$ ratio to two different 
isophotal levels, $I_1=I_{200}$ and $I_2=0.5\times I_{200}$, where
$I_{200}$ is the average luminosity density at the virial radius
of the supercluster core.

\section{Discussion and results}
\subsection{Supercluster constituents and extent}
Using early-type galaxies we have mapped SCL2243 over four square 
degrees. Its extent in East-West is $45\,h_{70}^{-1}$ Mpc,
whereas it appears more compact North-South ($15\,h_{70}^{-1}$ Mpc).  
If future spectroscopic observations confirm further objects from
Table \ref{clustertable}, the latter figure could increase to
$\sim 30\,h_{70}^{-1}$ Mpc. The redshift range of
$0.435-0.456$ suggests an extension of about 50 Mpc along the line 
of sight. This makes SCL2243 the largest supercluster known at
intermediate redshifts (Table \ref{knownsuperclusters}). Only
J1000+0231 at $z\sim0.7$ in the COSMOS field has an extent
of 400 Mpc along the line sight \citep{gcf07}, but one may argue that 
this is due to a projection of three unconnected smaller structures.

The supercluster centre is formed by a massive galaxy cluster,
SCL2243-A \citep[MACS J2243][]{eem10}, with
$r_{200}=2.06\pm0.13\,h_{70}^{-1}$ Mpc,
$N_{200}=150\pm30$ and $M_{200}=(1.54\pm0.29)\times10^{15}\,M_\odot$.
It shows significant sub-structuring, several strongly lensed galaxies
and is not in dynamic equilibrium (see Fig. \ref{scl2243_a} for an
image). More details about this spectacular cluster will be published
elsewhere (M. Brada{\v c}, priv. comm.).

The outer parts of SCL2243 are characterised by five individual 
clusters, SCL2243-B, H, M, N and L (Fig. 
\ref{scl2243_fullfield}). The latter two still need spectroscopic 
confirmation. The remaining objects listed in Table
\ref{clustertable} have very low concentration and do not exhibit
an obvious centre or a characteristic brightest cluster galaxy.
SCL2243-D, F, G, I (and K) form a $15-20$ Mpc long filament rich in
early-type galaxies, but therein they hardly represent more than just 
density variations. This pearl necklace appearance is typical for
inter-cluster filaments and commonly seen in simulations
\citep[][]{awj10}.

\begin{table*}
\begin{center}
\caption{Properties of the various supercluster components, integrated within
two different isophotal contours $I^{\rm iso}$: $1.0$ and $0.5$ times the average 
luminosity $I^{\rm iso}_{200}$ at the supercluster virial radius. The latter covers 
a $\sim2.6$ times larger area than the former. We list the integrated weak lensing 
mass, M/L ratio, colour, $\kappa$, and compare the
physical density to the critical density.}
\label{ml_filaments}
\large
\begin{tabular}{lcccccc}
\hline
\hline
\smallskip
 & $I^{\rm iso}/I^{\rm iso}_{200}$ & M [$10^{14}\,M_\odot$] & M/L & $\langle g-i\rangle$ &
$\langle \kappa \rangle$ & $\langle \rho/\rho_{\rm crit}\rangle$\\
\hline
MACS J2243         & $1.0$ & $15.4\pm2.9$  & $428\pm82$  & $2.06\pm0.01$ & $0.0436\pm0.0023$  &$200\pm37$\\
\hline
Filaments          & $1.0$ & $11.4\pm4.2$  & $409\pm153$ & $1.80\pm0.08$ & $0.0124\pm0.0023$ &$121\pm57$\\
Filaments          & $0.5$ & $15.3\pm10.1$ & $305\pm201$ & $1.75\pm0.08$ & $0.0065\pm0.0023$ &$39\pm29$\\
\hline
Filaments w/o$^a$  & $1.0$ & $5.2\pm3.5$   & $234\pm157$ & $1.79\pm0.08$ & $0.0068\pm0.0023$ &$59\pm46$\\
Filaments w/o$^a$  & $0.5$ & $6.6\pm9.1$   & $151\pm210$ & $1.75\pm0.08$ & $0.0031\pm0.0023$ &$15\pm25$\\
\hline
\end{tabular}
\\$^a$: Without the contribution of denser areas where the S/N of the
mass reconstruction is larger than 2.
\normalsize
\end{center}
\end{table*}

SCL2243-C (Fig. \ref{scl2243_c}) has $\sim$30 early-type galaxies,
but lacks a clear cluster centre. Yet it forms the second highest
weak lensing peak after MACS J2243. Actually, SCL2243-J has a stronger
lensing signal, but this is due to a foreground cluster at
$z=0.326$. SCL2243-C could be a filament oriented along the line of
sight. If it stretches over $10-20$ Mpc or more, then cosmic expansion
should lead to a characteristic broadening of the redshift
distribution distinguishing it from a virialised cluster.

The only typical galaxy cluster in addition to MACS J2243 inside the
CFHT field is SCL2243-H. It is about 0.2 mag redder than the filaments
(and 0.2 mag bluer than MACS J2243) and has two bright and merging
ellipticals at its core (Fig. \ref{scl2243_h}). We spectroscopically 
confirmed 15 galaxies in this cluster.

\subsection{Mass, light and colour of the cosmic web}
SCL2243 is an ideal laboratory to study the cosmic web. At a redshift
of $0.45$ and with an extent of two degrees it can easily be probed by
current wide field imagers $7-8$ magnitudes beyond the brightest
cluster galaxies. The intermediate cluster redshift makes it a good
target for weak lensing analyses.

\subsubsection{The supercluster centre: a normal but massive cluster}
For MACS J2243 we find excellent agreement for $r_{200}$ and $M_{200}$
between weak lensing and cluster scaling relations 
\citep[][]{jsw07,hsw09}. Its location at the centre of a supercluster
does not appear to alter its properties as compared to other clusters
in the MaxBCG sample. Its $M/L\sim428\pm82$ is in perfect agreement
with those measured by \cite{sjm09} for the MaxBCG cluster sample (see
their Fig. 9), with more massive clusters having higher $M/L$ ratios.
This is expected as very massive clusters arise preferentially at the
intersection of two or more cosmic filaments \citep{ckc05,awj10},
forming supercluster systems like SCL2243.

\subsubsection{Width and length of the filaments in SCL2243}
The global appearance of SCL2243, having a supermassive cluster at
the centre as well as several normal-sized clusters and filaments with
distinct density variations, is very characteristic \citep{awj10}. The
typical width of the three filaments is $(1.5-2.0)\,h_{70}^{-1}$ Mpc 
with local variations up to 3 Mpc. \cite{csb10} observed similar
values for a sample of high and low redshift filaments ($z\sim0.8$
and $z\sim0.1$, respectively) taken from the DEEP2 \citep{dfn03} and
SDSS surveys. Their width distribution 
peaks around $2\,h_{100}^{-1}$ Mpc and shows only little evolution 
with redshift (filaments tend to become more compact with
time, in agreement with $\Lambda$CDM). \cite{awj10} confirm this
picture in simulations, yielding a mean width of 
$2\,h_{70}^{-1}$ Mpc. \cite{gop10} predict a characteristic scale radius 
of $(1.0-1.2)\,h_{70}^{-1}$ Mpc for filaments less than 30 Mpc long,
increasing to $3$ Mpc for longer filaments.

With lengths of $(5-15)\,h_{70}^{-1}$ Mpc the filaments in SCL2243
are well matched by those of \cite{csb10} (see their Fig. 4 for a 
distribution of filament lengths). If further observations reveal
that filament DFGIK connects to SCL2243-M, then it would extend over
$\sim25$ Mpc. Only $2-3\%$ of the filaments found by \cite{csb10}
stretch over even longer distances.

\subsubsection{The colours of SCL2243}
The evolution of galaxies is a strong function of local density.
Star formation is already enhanced in filaments,
triggered by the increased number of interactions with neighbouring
galaxies. The consumption of gas and ram-pressure stripping quickly
turn blue galaxies into passive systems with colours similar to those
of red sequence galaxies 
\citep[see][and references therein]{qmb00,gzb04,por07,thk07,prp08,vzg08},
whereas morphological changes happen on significantly longer time
scales \citep{pef09,sws09}. 

For this reason one expects the density variation in a supercluster to
be mirrored in galaxy colours, an effect we see clearly for SCL2243 
(Fig. \ref{colormap}). The core of MACS J2243 is entirely
dominated by ellipticals with $\langle g-i \rangle\sim2.27$. At about
$0.25\times r_{200}$ or $\sim 500$ kpc galaxies start to become bluer,
a quasi-linear trend leading to $\langle g-i \rangle\sim1.90\pm0.05$
at the virial radius (2.1 Mpc). The filaments have on average a colour
of 1.8 mag, with a slight reddening towards their core by 0.05
mag. To some extent this central reddening is a consequence of
smoothing the data field with a 4\myarcmin0 wide kernel 
(about half the filament width), blending bluer field galaxies into
the filaments. Thus we do not consider colour variations transverse to
the filaments' main axis further. At the location where filaments AH
and AF merge and connect to the supercluster centre, we observe
increased reddening ($g-i\sim1.9$ mag) already 1.5 Mpc outside the
virial radius.

Outside this cluster infall region the filament colours stay
remarkably constant even over projected clustercentric distances as
large as 15 Mpc (filament DFGI). We do observe, however, an increase
in $g-i$ by about 0.05 mag where the filaments become thicker. These
locations appear correlated with higher surface mass density
(Fig. \ref{colormap}). While this is not statistically 
significant for most individual overdensities, the global picture is
more conclusive. 13 of the weak lensing peaks with $S/N>2$ are located
inside or on the edge of the filaments with $g-i>1.75$ mag,
whereas only four are found outside in areas with bluer colour
(the $g-i=1.75$ mag contour roughly divides the area of the field of 
view in half). The most significant peak in a blue area is caused by
a background cluster at $z\sim0.7$ ($\alpha=340.37;\;\delta=-09.32$;
note that we use galaxies with $0.40\leq z_{\rm phot}\leq 0.52$ only
to calculate colours). When turning to negative peaks,
i.e. under-dense regions, results are inverted. Eleven negative mass 
peaks are found in blue areas and only five within filaments. Outside
the evident skeleton of SCL2243 traced by early-type galaxies, the
mean colour of galaxies with $0.40\leq z_{\rm phot}\leq 0.52$ is
$1.60$ mag with little variation (0.05 mag).

\subsubsection{\label{filamentmass}Mass of the filaments}
A direct and statistically significant detection of an individual 
filament using weak lensing is very difficult and has not yet been
demonstrated (see Sect. \ref{filamentlensing}). Nevertheless, we
have shown that it is possible to constrain the mass and density of
the cosmic web by summing $\kappa$ over areas populated by
early-type galaxies. We integrated over two arbitrarily
chosen isophotes, $I_1=I_{200}\sim10^{9.9}\,L_\odot\,{\rm Mpc}^{-2}$
and $I_2=0.5\times\,I_{200}$, where $I_{200}$ is the mean luminosity
density at the virial radius of MACS J2243. $I_2$ and $I_1$ are 
represented by the outermost and second outermost contour lines, 
respectively, in Figs. \ref{scl2241_redsequ_specz_045}
and \ref{colormap}. Within $I_1$ only isolated overdensities in the 
filaments are selected. $I_2$ on the other hand encompasses a three
times larger area. Our measurement is limited to filaments AC, AH, and
AFDGI, and our results are shown in Table \ref{ml_filaments}.

Within $I_1$ we find a mass of $(1.1\pm0.4)\times10^{15}\,M_\odot$,
increasing to $(1.53\pm1.01)\times10^{15}\,M_\odot$ for $I_2$.
Hence these filaments contain about the same mass as MACS J2243
($M_{200}=(1.54\pm0.29)\times10^{15}\,M_\odot$). This may sound
surprising, but is predicted by $\Lambda$CDM models 
which identify filaments as the carriers of most of the mass in the
Universe, with clusters being a close second \citep{awj10}.
Similar results as for values were obtained for the Shapley
supercluster, where galaxies in the filaments contribute about twice
as much mass as those in clusters \citep{pqc06}. For comparison, 
simplistic linear theory yields that filaments and walls contain 
about 10 times as much mass as clusters \citep[][for an open $\Lambda=0$
Friedmann Universe]{dor70, awj10}.

The mass-to-light ratio for the SCL2243
filaments ($409\pm153$) within $I_1$ is indistinguishable from MACS
J2243 ($480\pm82$) and becomes smaller ($305\pm201$) when integrating
over the larger $I_2$ area. The increased error is caused by
integrating $\kappa$ over areas further away from the filament 
cores where the signal is diminished.

\subsubsection{Density of the filaments}
We find $\langle \kappa \rangle=0.0124\pm0.0023$ for the denser
filament regions within $I_1$, decreasing to $0.0065\pm0.0023$ when
integrating over $I_2$. This is in agreement with theoretical
expectations of $\langle\kappa\rangle\sim0.01$ \citep[][]{mkm10}.
Errors are fully dominated by the uncertainty of the mass sheet
degeneracy (Sect. \ref{massrec}). Making assumptions about the
three-dimensional shape of the filaments and embedded condensations,
we can de-project $\kappa$ and estimate the physical density. However,
there are some uncertainties involved.

First, our knowledge about foreground and background matter is
mostly limited to three isolated NTT/EMMI pointings. SCL2243-J
was excluded from our analysis as the spectra revealed that it is
located behind a foreground cluster (Table \ref{EMMI_POS}). The
spectra for SCL2243-H show a thin foreground sheet at $z\sim0.256$,
but the galaxies associated with it are field spirals and do not form
a spatially concentrated structure. Their lensing impact is
negligible. For SCL2243-G and -I there is no evidence for other
structures along the line of sight. Scanning the CFHT field for
concentrations of red sequence galaxies at other redshifts did
not reveal structures projected on top of SCL2243 that could cause a
measurable lensing signal. We do see very high redshift clusters at
$z\gtrsim 0.8$ across the field, too distant to affect our
results. Other structures, such as the cluster at
$\alpha=340.37;\;\delta=-09.32$ ($z\sim0.7$,
Fig. \ref{scl2241_redsequ_specz_045}) or concentrations of early-type 
galaxies at lower photometric redshifts are not projected onto cluster 
filaments. Hence we assume that contamination along the line of
sight does not bias our mass and density estimates significantly.

The second hurdle in estimating the physical density of a filament is
the determination of its volume. The density contrast is much lower
than that of clusters, for which reliable size estimators are
available (Sect. \ref{n200}). Different measures exist for a
filaments' characteristic radius, based on the galaxy distribution,
the typical orbit time of a galaxy in the filament, or the shape of
its density profile \citep{ckc05,gop10}. To cover these uncertainties
we probe the density within the $I_1$ and $I_2$ contours allowing for
an additional $\pm0.2$ Mpc error. Within $I_1$ the filaments decompose
mostly into individual isolated overdensities, for which we assume
spherical shapes. The average density in units of $\rho_{\rm crit}$
is $121\pm57$, about half the mean density of MACS J2243 within
the same isophote ($I_1$ is the average luminosity at $r_{200}$ of
MACS J2243). Within $I_2$, the assumption of a spherical shape is
invalid and we adopt a cylindrical model. The redshifts of SCL2243-F
and G ($z=0.437$) and MACS J2243 ($z=0.447$) indicate that this
filament is inclined by 60 degrees with respect to the celestial
plane. SCL2243-H on the other hand is at the same redshift as MACS
J2243. Thus we assume some arbitrary mean inclination of 30 degrees,
resulting in $\rho/\rho_{\rm crit}=39\pm29$. The error bars include
the uncertainties of the volume and of the mass sheet degeneracy, as
well as the noise of the mass reconstruction itself. These are the
mean filament densities including local overdensities.

By integrating within the same isophotes, but by excluding all
regions where $S/N(\kappa)>2$, we measure the density of the stripped
filaments, i.e. the parts without local overdensities. Within $I_1$ 
($I_2$) we have $\rho/\rho_{\rm crit}=27\pm46$ ($7\pm25$). Both
results are positive but consistent with zero. In case of $I_1$, the
uncertainty of the mass sheet degeneracy prevents a detection.

For comparison, \cite{ckc05} found the filament density 
to vary between a few and 25 times the mean density of the 
Universe in $\Lambda$CDM simulations. With $\Omega_{\rm m}=0.3$ this
is consistent with our results (scaled in terms of $\rho_{\rm crit}$).

\section{Summary}
\subsection{Overview}
In one of the fields of our weak lensing cluster survey \citep{seh07}
we made several detections. Three of them (objects \#141, 144 and 147)
are associated with significant overdensities of galaxies with
singular SDSS redshifts of $z\sim0.45$, indicating a supercluster of
galaxies. This would be the first supercluster discovered using weak
gravitational lensing methods only.

Our spectroscopic follow-up with NTT/EMMI confirmed the SDSS redshift
of \#141 (SCL2243-J, Fig. \ref{scl2243_j}) and revealed two
more galaxies at the same redshift. The bulk of the other 31 galaxies
in this area belong to a foreground cluster at $z=0.326$. We
cannot rule out that there are more galaxies at $z=0.456$, as our
spectroscopic coverage is shallow ($i\sim22$) and only half
complete at this depth. Object \#144 (SCL2243-H, Fig. \ref{scl2243_h})
was confirmed with 15 galaxies at $z\sim0.447$, and \#147
(SCL2243-G and I, Fig. \ref{scl2243_g}) has 19 confirmed member
galaxies at $z=0.436$. Depth and completeness are the same as for
\#141.

To map the larger extent of SCL2243 and to identify its centre we
observed one square degree with CFHT/Megaprime in $ugriz$ filters. 
In addition, twelve pointings were taken in $gri$ with the 2.5m
INT/WFC (La Palma, Spain) increasing the sky coverage to four square
degrees. By selecting early-type galaxies at $z\sim0.45$ we detected
14 potential member clusters or overdensities 
(Fig. \ref{scl2243_fullfield}, Table \ref{clustertable}), eight of
which have positive spectroscopic confirmation of at least the
brightest galaxy. SCL2243-B, H, L, M and N are typical galaxy
clusters. Further spectroscopic confirmation is required. In
particular the region between clusters K, M and N has a significantly 
increased density of early type galaxies as compared to the remaining 
inter-cluster areas (Fig. \ref{scl2243_fullfield}). This should be
verified by future field spectroscopy.

Using early-type galaxies we traced the supercluster skeleton of 
SCL2243 over clustercentric distances of $30\,h_{70}^{-1}$ Mpc 
(Fig. \ref{scl2243_fullfield}). We identified three large filaments
and mapped the colours of the central square degree from high to
low density regions. We determined the filaments' masses by
integrating the reconstructed surface density over areas with enhanced
$i$-band luminosity, respectively red $g-i$ colours.

SCL2243 is the third supercluster (with more than two member
clusters) next to A901/902 and MS0302 that has been studied with 
weak lensing. Cores of other superclusters (usually supermassive
clusters) at intermediate redshift have been studied with lensing
before, but this is the first time that such methods have been applied
to the cosmic web in superclusters.

\subsection{Main results}
\begin{itemize}
\item{We have made the first discovery of a supercluster, SCL2243,
  using the shear-selection method. Interestingly, the initial
  detections were made on smaller clusters located in the supercluster
  filamentary network, as the supercluster centre was outside the
  initial field of view.}
\item{SCL2243 extends over 
  $45\times15\times50\,h_{70}^{-1}$ Mpc (length, width, depth) and is
  one of the largest superclusters studied at intermediate
  redshifts. We identified 14 member clusters and
  overdensities, as well as three filaments extending over  
  $(5-15)\,h_{70}^{-1}$ Mpc and possibly beyond. Almost all clusters
  or overdensities found in the CFHT data are embedded in
  filaments. We expect that better data on the surrounding INT
  fields will confirm this picture for the remaining
  clusters.}
\item{The supercluster centre, SCL2243-A, was found at the edge
  of the CFHT field. With $M_{200}=(1.3-1.5)\times10^{15}M_\odot$ and
  $M/L=428\pm82$ it is fully consistent with similar clusters in the
  MaxBCG sample. SCL2243-A is also contained in the massive cluster
  survey \cite[][]{eeh01} and referred to as MACS
  J2243-0935. There is very good agreement between cluster scaling
  relations and weak lensing based mass and size estimates
  ($r_{200}=2.06\pm0.13\,h_{70}^{-1}$ Mpc).}
\item{We observe a significant correlation between the
  surface mass density $\kappa$ and the supercluster skeleton as  
  outlined by its $i$-band luminosity and $g-i$ colours. $\kappa$
  is statistically insignificantly enhanced above zero
  for most areas, yet a global trend is visible. Motivated by
  this, we integrated $\kappa$ over the filaments and
  obtained for the first time good weak lensing constraints for the
  cosmic web. We find a total filament mass of
  $(1.53\pm10.1)\times10^{15}\,M_\odot$, similar to 
  $M_{200}=(1.54\pm0.29)\times10^{15}M_\odot$ for MACS 
  J2243. More mass is located in another filament which we
  predict at the South-East of MACS J2243. The high filament mass is
  in agreement with $\Lambda$CDM models where most of the mass of the 
  Universe is in filaments, with galaxy clusters being second.}
\item{The measured weak lensing surface mass density of the filaments
  is $\kappa\sim 0.003-0.012$, consistent with
  theoretical expectations ($\kappa\sim0.01$). Uncertainties in the
  determination of the mass sheet degeneracy become the limiting 
  factor and prevent a positive detection when we aim at the naked
  parts of the filament, i.e. those regions without overdensities.
  Depending on de-projection, these surface mass densities
  translate to physical densities of 
  $\rho/\rho_{\rm crit}\sim10-100$.} 
\item{The $M/L$ ratio of the filaments evaluates to $409\pm153$, very
  similar to that of MACS J2243 ($428\pm82$), integrated within the
  same isophotal contour.}
\item{The density field in SCL2243 shows large variations which are
  typical for superclusters, ranging from supermassive clusters down
  to low contrast filaments. Correspondingly, we expect galaxy colours
  to change depending on their location. The
  supercluster core has $\langle g-i\rangle =2.27$ mag, dropping to
  1.90 mag at $r_{200}$. For the filaments we observe a constant
  colour of 1.80 mag, independent of clustercentric distance. Only in
  the cluster infall region (out to 1.5 Mpc outside $r_{200}$) do the
  filaments become noticeably redder, having the same average colour
  as the supercluster centre at its virial radius. Small colour
  variations of $0.05$ mag amplitude are observed along the filaments,
  coinciding with local overdensities or one of the member clusters.}
\end{itemize}

SCL2243 and its massive filamentary network form a textbook 
supercluster. Large-scale spectroscopic coverage is required to
disentangle its real 3D shape and to obtain independent mass
estimates. With very deep wide-field imaging using 8m class 
telescopes it should be possible to make a significantly improved weak
lensing study of the filamentary network in SCL2243. Our new approach
of directly probing the cosmic web with weak lensing should facilitate
future lensing-based mass estimates of inter-cluster filaments. 

\begin{acknowledgements}
MS thanks Catherine Heymans, Karianne Holhjem, Lindsay King, Emilio 
Pastor Mira, Peter Schneider, and the anonymous referee for
discussions and suggestions  
which improved this paper. The authors wish to recognise and
acknowledge the very significant cultural role and reverence that the
summit of Mauna Kea has always had within the indigenous Hawaiian
community. We are most fortunate to have the opportunity to conduct
observations from this mountain. MS acknowledges support by the German
Ministry for Science and Education (BMBF) through DESY under the
project 05AV5PDA/3 and the Deutsche Forschungsgemeinschaft (DFG) in
the frame of the Schwerpunktprogramm SPP 1177 `Galaxy Evolution'. HH
is supported by the DUEL-RTN, MRTN-CT-2006-036133, the Marie Curie
individual fellowship PIOF-GA-2009-252760, and a CITA National
Fellowship. TE is supported by the German Ministry for Science and
Education (BMBF) through the DESY project "GAVO III" and by the
Deutsche Forschungsgemeinschaft through project ER 327/3-1 and the
Transregional Collaborative Research Centre TR 33 $-$ "The Dark
Universe".

Author contributions: MS obtained the CFHT, INT and NTT data, 
reduced the INT images and NTT spectra, and performed all scientific 
analyses. HH derived the photometric redshifts based on CFHT data.
KK developed the PSF Gaussianisation code used in the photometry that
led to the photometric redshifts. TE stacked the {\tt Elixir} 
pre-processed CFHT $ugriz$ images. 

This research has made use of the VizieR catalogue 
access tool, CDS, Strasbourg, France. Funding for the SDSS and SDSS-II
has been provided by the Alfred P. Sloan Foundation, the Participating
Institutions, the National Science Foundation, the U.S. Department of
Energy, the National Aeronautics and Space Administration, the
Japanese Monbukagakusho, the Max Planck Society, and the Higher
Education Funding Council for England. The SDSS Web Site is
http://www.sdss.org/. The SDSS is managed by the Astrophysical
Research Consortium for the Participating Institutions. The
Participating Institutions are the American Museum of Natural History,
Astrophysical Institute Potsdam, University of Basel, University of
Cambridge, Case Western Reserve University, University of Chicago,
Drexel University, Fermilab, the Institute for Advanced Study, the
Japan Participation Group, Johns Hopkins University, the Joint
Institute for Nuclear Astrophysics, the Kavli Institute for Particle
Astrophysics and Cosmology, the Korean Scientist Group, the Chinese
Academy of Sciences (LAMOST), Los Alamos National Laboratory, the
Max-Planck-Institute for Astronomy (MPIA), the Max-Planck-Institute
for Astrophysics (MPA), New Mexico State University, Ohio State
University, University of Pittsburgh, University of Portsmouth,
Princeton University, the United States Naval Observatory, and the
University of Washington.
\end{acknowledgements}

\bibliography{mybib}

\begin{thebibliography}{67}
\expandafter\ifx\csname natexlab\endcsname\relax\def\natexlab#1{#1}\fi

\bibitem[{{Adelman-McCarthy} {et~al.}(2008){Adelman-McCarthy}, {Ag{\"u}eros},
  {Allam}, {Allende Prieto}, {Anderson}, {Anderson}, {et~al.}}]{aaa08}
{Adelman-McCarthy}, J.~K., {Ag{\"u}eros}, M.~A., {Allam}, S.~S., {et~al.} 2008,
  \apjs, 175, 297

\bibitem[{{Arag{\'o}n-Calvo} {et~al.}(2010){Arag{\'o}n-Calvo}, {van de
  Weygaert}, \& {Jones}}]{awj10}
{Arag{\'o}n-Calvo}, M.~A., {van de Weygaert}, R., \& {Jones}, B.~J.~T. 2010,
  \mnras, 408, 2163

\bibitem[{{Araya-Melo} {et~al.}(2009){Araya-Melo}, {Reisenegger}, {Meza}, {van
  de Weygaert}, {D{\"u}nner}, \& {Quintana}}]{arm09}
{Araya-Melo}, P.~A., {Reisenegger}, A., {Meza}, A., {et~al.} 2009, \mnras, 399,
  97

\bibitem[{Bartelmann \& Schneider(2001)}]{bas01}
Bartelmann, M. \& Schneider, P. 2001, Phys. Rep., 340, 291

\bibitem[{{Bertin}(2006)}]{ber06}
{Bertin}, E. 2006, in Astronomical Society of the Pacific Conference Series,
  Vol. 351, Astronomical Data Analysis Software and Systems XV, ed.
  C.~{Gabriel}, C.~{Arviset}, D.~{Ponz}, \& S.~{Enrique}, 112--+

\bibitem[{Bertin \& Arnouts(1996)}]{bea96}
Bertin, E. \& Arnouts, S. 1996, A\&AS, 117, 393

\bibitem[{{Bertin} {et~al.}(2002){Bertin}, {Mellier}, {Radovich}, {Missonnier},
  {Didelon}, \& {Morin}}]{ber02}
{Bertin}, E., {Mellier}, Y., {Radovich}, M., {et~al.} 2002, in Astronomical
  Society of the Pacific Conference Series, Vol. 281, Astronomical Data
  Analysis Software and Systems XI, ed. {D.~A.~Bohlender, D.~Durand, \&
  T.~H.~Handley}, 228--+

\bibitem[{{Blanton} \& {Roweis}(2007)}]{blr07}
{Blanton}, M.~R. \& {Roweis}, S. 2007, \aj, 133, 734

\bibitem[{{Bond} {et~al.}(1996){Bond}, {Kofman}, \& {Pogosyan}}]{bkp96}
{Bond}, J.~R., {Kofman}, L., \& {Pogosyan}, D. 1996, \nat, 380, 603

\bibitem[{{Choi} {et~al.}(2010){Choi}, {Bond}, {Strauss}, {Coil}, {Davis}, \&
  {Willmer}}]{csb10}
{Choi}, E., {Bond}, N.~A., {Strauss}, M.~A., {et~al.} 2010, \mnras, 406, 320

\bibitem[{{Clowe} {et~al.}(2006){Clowe}, {Brada{\v c}}, {Gonzalez},
  {Markevitch}, {Randall}, {Jones}, \& {Zaritsky}}]{cbg06}
{Clowe}, D., {Brada{\v c}}, M., {Gonzalez}, A.~H., {et~al.} 2006, \apjl, 648,
  L109

\bibitem[{{Colberg} {et~al.}(2005){Colberg}, {Krughoff}, \& {Connolly}}]{ckc05}
{Colberg}, J.~M., {Krughoff}, K.~S., \& {Connolly}, A.~J. 2005, \mnras, 359,
  272

\bibitem[{{Colless} {et~al.}(2001){Colless}, {Dalton}, {Maddox}, {Sutherland},
  {Norberg}, {Cole}, {Bland-Hawthorn}, {Bridges}, {Cannon}, {Collins}, {Couch},
  {Cross}, {Deeley}, {De Propris}, {Driver}, {Efstathiou}, {Ellis}, {Frenk},
  {Glazebrook}, {Jackson}, {Lahav}, {Lewis}, {Lumsden}, {Madgwick}, {Peacock},
  {Peterson}, {Price}, {Seaborne}, \& {Taylor}}]{colless01}
{Colless}, M., {Dalton}, G., {Maddox}, S., {et~al.} 2001, \mnras, 328, 1039

\bibitem[{{Dahle}(2007)}]{dah07}
{Dahle}, H. 2007, ArXiv Astrophysics e-prints

\bibitem[{{Davis} {et~al.}(2003){Davis}, {Faber}, {Newman}, {Phillips},
  {Ellis}, {Steidel}, {Conselice}, {Coil}, {Finkbeiner}, {Koo}, {Guhathakurta},
  {Weiner}, {Schiavon}, {Willmer}, {Kaiser}, {Luppino}, {Wirth}, {Connolly},
  {Eisenhardt}, {Cooper}, \& {Gerke}}]{dfn03}
{Davis}, M., {Faber}, S.~M., {Newman}, J., {et~al.} 2003, in Society of
  Photo-Optical Instrumentation Engineers (SPIE) Conference Series, Vol. 4834,
  Society of Photo-Optical Instrumentation Engineers (SPIE) Conference Series,
  ed. {P.~Guhathakurta}, 161--172

\bibitem[{{Deb} {et~al.}(2010){Deb}, {Goldberg}, {Heymans}, \&
  {Morandi}}]{dgh10}
{Deb}, S., {Goldberg}, D.~M., {Heymans}, C., \& {Morandi}, A. 2010, \apj

\bibitem[{{Dietrich} {et~al.}(2007){Dietrich}, {Erben}, {Lamer}, {Schneider},
  {Schwope}, {Hartlap}, \& {Maturi}}]{del07}
{Dietrich}, J.~P., {Erben}, T., {Lamer}, G., {et~al.} 2007, \aap, 470, 821

\bibitem[{{Dietrich} {et~al.}(2005){Dietrich}, {Schneider}, {Clowe},
  {Romano-D{\'{\i}}az}, \& {Kerp}}]{dsc05}
{Dietrich}, J.~P., {Schneider}, P., {Clowe}, D., {Romano-D{\'{\i}}az}, E., \&
  {Kerp}, J. 2005, \aap, 440, 453

\bibitem[{{Doroshkevich}(1970)}]{dor70}
{Doroshkevich}, A.~G. 1970, Astrophysics, 6, 320

\bibitem[{{Ebeling} {et~al.}(2001){Ebeling}, {Edge}, \& {Henry}}]{eeh01}
{Ebeling}, H., {Edge}, A.~C., \& {Henry}, J.~P. 2001, \apj, 553, 668

\bibitem[{{Ebeling} {et~al.}(2010){Ebeling}, {Edge}, {Mantz}, {Barrett},
  {Henry}, {Ma}, \& {van Speybroeck}}]{eem10}
{Ebeling}, H., {Edge}, A.~C., {Mantz}, A., {et~al.} 2010, \mnras, 407, 83

\bibitem[{{Einasto} {et~al.}(2007{\natexlab{a}}){Einasto}, {Einasto}, {Saar},
  {Tago}, {Liivam{\"a}gi}, {J{\~o}eveer}, {Suhhonenko}, {H{\"u}tsi},
  {Jaaniste}, {Hein{\"a}m{\"a}ki}, {M{\"u}ller}, {Knebe}, \& {Tucker}}]{ees07}
{Einasto}, J., {Einasto}, M., {Saar}, E., {et~al.} 2007{\natexlab{a}}, \aap,
  462, 397

\bibitem[{{Einasto} {et~al.}(2007{\natexlab{b}}){Einasto}, {Einasto}, {Tago},
  {Saar}, {H{\"u}tsi}, {J{\~o}eveer}, {Liivam{\"a}gi}, {Suhhonenko},
  {Jaaniste}, {Hein{\"a}m{\"a}ki}, {M{\"u}ller}, {Knebe}, \& {Tucker}}]{eet07}
{Einasto}, J., {Einasto}, M., {Tago}, E., {et~al.} 2007{\natexlab{b}}, \aap,
  462, 811

\bibitem[{{Einasto} {et~al.}(2007{\natexlab{c}}){Einasto}, {Saar},
  {Liivam{\"a}gi}, {Einasto}, {Tago}, {Mart{\'{\i}}nez}, {Starck},
  {M{\"u}ller}, {Hein{\"a}m{\"a}ki}, {Nurmi}, {Gramann}, \&
  {H{\"u}tsi}}]{esl07}
{Einasto}, M., {Saar}, E., {Liivam{\"a}gi}, L.~J., {et~al.} 2007{\natexlab{c}},
  \aap, 476, 697

\bibitem[{{Einasto} {et~al.}(2010){Einasto}, {Tago}, {Saar}, {Nurmi},
  {Enkvist}, {Einasto}, {Hein{\"a}m{\"a}ki}, {Liivam{\"a}gi}, {Tempel},
  {Einasto}, {Mart{\'{\i}}nez}, {Vennik}, \& {Pihajoki}}]{ets10}
{Einasto}, M., {Tago}, E., {Saar}, E., {et~al.} 2010, \aap, 522, A92+

\bibitem[{{Erben} {et~al.}(2009){Erben}, {Hildebrandt}, {Lerchster}, {Hudelot},
  {Benjamin}, {van Waerbeke}, {Schrabback}, {Brimioulle}, {Cordes}, {Dietrich},
  {Holhjem}, {Schirmer}, \& {Schneider}}]{ehl09}
{Erben}, T., {Hildebrandt}, H., {Lerchster}, M., {et~al.} 2009, \aap, 493, 1197

\bibitem[{Erben {et~al.}(2005)Erben, Schirmer, Dietrich, {et~al.}}]{esd05}
Erben, T., Schirmer, M., Dietrich, J., {et~al.} 2005, AN, 326, 432

\bibitem[{Erben {et~al.}(2001)Erben, van Waerbeke, Bertin, Mellier, \&
  Schneider}]{ewb01}
Erben, T., van Waerbeke, L., Bertin, E., Mellier, Y., \& Schneider, P. 2001,
  A\&A, 366, 717

\bibitem[{{Gal} {et~al.}(2008){Gal}, {Lemaux}, {Lubin}, {Kocevski}, \&
  {Squires}}]{gll08}
{Gal}, R.~R., {Lemaux}, B.~C., {Lubin}, L.~M., {Kocevski}, D., \& {Squires},
  G.~K. 2008, \apj, 684, 933

\bibitem[{{Gavazzi} {et~al.}(2004){Gavazzi}, {Mellier}, {Fort}, {Cuillandre},
  \& {Dantel-Fort}}]{gmf04}
{Gavazzi}, R., {Mellier}, Y., {Fort}, B., {Cuillandre}, J., \& {Dantel-Fort},
  M. 2004, \aap, 422, 407

\bibitem[{{Gerken} {et~al.}(2004){Gerken}, {Ziegler}, {Balogh}, {Gilbank},
  {Fritz}, \& {J{\"a}ger}}]{gzb04}
{Gerken}, B., {Ziegler}, B., {Balogh}, M., {et~al.} 2004, \aap, 421, 59

\bibitem[{{Gonz{\'a}lez} \& {Padilla}(2010)}]{gop10}
{Gonz{\'a}lez}, R.~E. \& {Padilla}, N.~D. 2010, \mnras, 407, 1449

\bibitem[{{Gramann} \& {Suhhonenko}(2002)}]{grs02}
{Gramann}, M. \& {Suhhonenko}, I. 2002, \mnras, 337, 1417

\bibitem[{{Gray} {et~al.}(2002){Gray}, {Taylor}, {Meisenheimer}, {Dye}, {Wolf},
  \& {Thommes}}]{gtm02}
{Gray}, M.~E., {Taylor}, A.~N., {Meisenheimer}, K., {et~al.} 2002, \apj, 568,
  141

\bibitem[{{Guzzo} {et~al.}(2007){Guzzo}, {Cassata}, {Finoguenov}, {Massey},
  {Scoville}, {Capak}, {Ellis}, {Mobasher}, {Taniguchi}, {Thompson}, {Ajiki},
  {Aussel}, {B{\"o}hringer}, {Brusa}, {Calzetti}, {Comastri}, {Franceschini},
  {Hasinger}, {Kasliwal}, {Kitzbichler}, {Kneib}, {Koekemoer}, {Leauthaud},
  {McCracken}, {Murayama}, {Nagao}, {Rhodes}, {Sanders}, {Sasaki}, {Shioya},
  {Tasca}, \& {Taylor}}]{gcf07}
{Guzzo}, L., {Cassata}, P., {Finoguenov}, A., {et~al.} 2007, \apjs, 172, 254

\bibitem[{{Hansen} {et~al.}(2005){Hansen}, {McKay}, {Wechsler}, {Annis},
  {Sheldon}, \& {Kimball}}]{hmw05}
{Hansen}, S.~M., {McKay}, T.~A., {Wechsler}, R.~H., {et~al.} 2005, \apj, 633,
  122

\bibitem[{{Hansen} {et~al.}(2009){Hansen}, {Sheldon}, {Wechsler}, \&
  {Koester}}]{hsw09}
{Hansen}, S.~M., {Sheldon}, E.~S., {Wechsler}, R.~H., \& {Koester}, B.~P. 2009,
  \apj, 699, 1333

\bibitem[{{Heymans} {et~al.}(2008){Heymans}, {Gray}, {Peng}, {van Waerbeke},
  {Bell}, {Wolf}, {Bacon}, {Balogh}, {Barazza}, {Barden}, {B{\"o}hm},
  {Caldwell}, {H{\"a}u{\ss}ler}, {Jahnke}, {Jogee}, {van Kampen}, {Lane},
  {McIntosh}, {Meisenheimer}, {Mellier}, {S{\'a}nchez}, {Taylor}, {Wisotzki},
  \& {Zheng}}]{hgp08}
{Heymans}, C., {Gray}, M.~E., {Peng}, C.~Y., {et~al.} 2008, \mnras, 385, 1431

\bibitem[{{Hildebrandt} {et~al.}(2009){Hildebrandt}, {Pielorz}, {Erben}, {van
  Waerbeke}, {Simon}, \& {Capak}}]{hpe09}
{Hildebrandt}, H., {Pielorz}, J., {Erben}, T., {et~al.} 2009, \aap, 498, 725

\bibitem[{{Hoekstra} {et~al.}(2001){Hoekstra}, {Franx}, {Kuijken}, {Carlberg},
  {Yee}, {Lin}, {Morris}, {Hall}, {Patton}, {Sawicki}, \& {Wirth}}]{hfk01}
{Hoekstra}, H., {Franx}, M., {Kuijken}, K., {et~al.} 2001, \apjl, 548, L5

\bibitem[{Hoekstra {et~al.}(1998)Hoekstra, Franx, Kuijken, \& Squires}]{hfk98}
Hoekstra, H., Franx, M., Kuijken, K., \& Squires, G. 1998, ApJ, 504, 636

\bibitem[{{Johnston} {et~al.}(2007){Johnston}, {Sheldon}, {Wechsler}, {Rozo},
  {Koester}, {Frieman}, {McKay}, {Evrard}, {Becker}, \& {Annis}}]{jsw07}
{Johnston}, D.~E., {Sheldon}, E.~S., {Wechsler}, R.~H., {et~al.} 2007,
  astro-ph/0709.1159

\bibitem[{Kaiser \& Squires(1993)}]{kas93}
Kaiser, N. \& Squires, G. 1993, ApJ, 404, 441

\bibitem[{Kaiser {et~al.}(1995)Kaiser, Squires, \& Broadhurst}]{ksb95}
Kaiser, N., Squires, G., \& Broadhurst, T. 1995, ApJ, 449, 460

\bibitem[{{Kaiser} {et~al.}(1998){Kaiser}, {Wilson}, {Luppino}, {Kofman},
  {Gioia}, {Metzger}, \& {Dahle}}]{kwl98}
{Kaiser}, N., {Wilson}, G., {Luppino}, G., {et~al.} 1998, ArXiv Astrophysics
  e-prints

\bibitem[{{Luppino} \& {Kaiser}(1997)}]{luk97}
{Luppino}, G.~A. \& {Kaiser}, N. 1997, \apj, 475, 20

\bibitem[{Magnier \& Cuillandre(2004)}]{mac04}
Magnier, G. \& Cuillandre, J.-C. 2004, \pasp, 116, 449

\bibitem[{{Mead} {et~al.}(2010){Mead}, {King}, \& {McCarthy}}]{mkm10}
{Mead}, J.~M.~G., {King}, L.~J., \& {McCarthy}, I.~G. 2010, \mnras, 401, 2257

\bibitem[{{Merluzzi} {et~al.}(2010){Merluzzi}, {Mercurio}, {Haines}, {Smith},
  {Busarello}, \& {Lucey}}]{mmh10}
{Merluzzi}, P., {Mercurio}, A., {Haines}, C.~P., {et~al.} 2010, \mnras, 402,
  753

\bibitem[{Micol {et~al.}(2004)Micol, Pierfederici, Benvenuti, {et~al.}}]{mpb04}
Micol, A., Pierfederici, F., Benvenuti, P., {et~al.} 2004, ASPC, 314, 197

\bibitem[{{Parker} {et~al.}(2005){Parker}, {Hudson}, {Carlberg}, \&
  {Hoekstra}}]{phc05}
{Parker}, L.~C., {Hudson}, M.~J., {Carlberg}, R.~G., \& {Hoekstra}, H. 2005,
  \apj, 634, 806

\bibitem[{{Parry} {et~al.}(2009){Parry}, {Eke}, \& {Frenk}}]{pef09}
{Parry}, O.~H., {Eke}, V.~R., \& {Frenk}, C.~S. 2009, \mnras, 396, 1972

\bibitem[{{Porter} \& {Raychaudhury}(2007)}]{por07}
{Porter}, S.~C. \& {Raychaudhury}, S. 2007, \mnras, 375, 1409

\bibitem[{{Porter} {et~al.}(2008){Porter}, {Raychaudhury}, {Pimbblet}, \&
  {Drinkwater}}]{prp08}
{Porter}, S.~C., {Raychaudhury}, S., {Pimbblet}, K.~A., \& {Drinkwater}, M.~J.
  2008, \mnras, 388, 1152

\bibitem[{{Proust} {et~al.}(2006){Proust}, {Quintana}, {Carrasco},
  {Reisenegger}, {Slezak}, {Muriel}, {D{\"u}nner}, {Sodr{\'e}}, {Drinkwater},
  {Parker}, \& {Ragone}}]{pqc06}
{Proust}, D., {Quintana}, H., {Carrasco}, E.~R., {et~al.} 2006, \aap, 447, 133

\bibitem[{{Quilis} {et~al.}(2000){Quilis}, {Moore}, \& {Bower}}]{qmb00}
{Quilis}, V., {Moore}, B., \& {Bower}, R. 2000, Science, 288, 1617

\bibitem[{{Rykoff} {et~al.}(2008){Rykoff}, {McKay}, {Becker}, {Evrard},
  {Johnston}, {Koester}, {Rozo}, {Sheldon}, \& {Wechsler}}]{rmb08}
{Rykoff}, E.~S., {McKay}, T.~A., {Becker}, M.~R., {et~al.} 2008, \apj, 675,
  1106

\bibitem[{{Scannapieco} {et~al.}(2009){Scannapieco}, {White}, {Springel}, \&
  {Tissera}}]{sws09}
{Scannapieco}, C., {White}, S.~D.~M., {Springel}, V., \& {Tissera}, P.~B. 2009,
  \mnras, 396, 696

\bibitem[{Schechter(1976)}]{sch76}
Schechter, P. 1976, ApJ, 203, 297

\bibitem[{{Schirmer} {et~al.}(2007){Schirmer}, {Erben}, {Hetterscheidt}, \&
  {Schneider}}]{seh07}
{Schirmer}, M., {Erben}, T., {Hetterscheidt}, M., \& {Schneider}, P. 2007,
  \aap, 462, 875

\bibitem[{Seitz \& Schneider(2001)}]{ses01}
Seitz, S. \& Schneider, P. 2001, A\&A, 374, 740

\bibitem[{{Sheldon} {et~al.}(2004){Sheldon}, {Johnston}, {Frieman}, {Scranton},
  {McKay}, {Connolly}, {Budav{\'a}ri}, {Zehavi}, {Bahcall}, {Brinkmann}, \&
  {Fukugita}}]{sjf04}
{Sheldon}, E.~S., {Johnston}, D.~E., {Frieman}, J.~A., {et~al.} 2004, \aj, 127,
  2544

\bibitem[{{Sheldon} {et~al.}(2009){Sheldon}, {Johnston}, {Masjedi}, {McKay},
  {Blanton}, {Scranton}, {Wechsler}, {Koester}, {Hansen}, {Frieman}, \&
  {Annis}}]{sjm09}
{Sheldon}, E.~S., {Johnston}, D.~E., {Masjedi}, M., {et~al.} 2009, \apj, 703,
  2232

\bibitem[{{Swinbank} {et~al.}(2007){Swinbank}, {Edge}, {Smail}, {Stott},
  {Bremer}, {Sato}, {van Breukelen}, {Jarvis}, {Waddington}, {Clewley},
  {Bergeron}, {Cotter}, {Dye}, {Geach}, {Gonzalez-Solares}, {Hirst}, {Ivison},
  {Rawlings}, {Simpson}, {Smith}, {Verma}, \& {Yamada}}]{swinbank07}
{Swinbank}, A.~M., {Edge}, A.~C., {Smail}, I., {et~al.} 2007, \mnras, 379, 1343

\bibitem[{{Tanaka} {et~al.}(2007){Tanaka}, {Hoshi}, {Kodama}, \&
  {Kashikawa}}]{thk07}
{Tanaka}, M., {Hoshi}, T., {Kodama}, T., \& {Kashikawa}, N. 2007, \mnras, 379,
  1546

\bibitem[{{Verdugo} {et~al.}(2008){Verdugo}, {Ziegler}, \& {Gerken}}]{vzg08}
{Verdugo}, M., {Ziegler}, B.~L., \& {Gerken}, B. 2008, \aap, 486, 9

\bibitem[{{Woudt}(2009)}]{wou09}
{Woudt}, P.~A. 2009, in Panoramic Radio Astronomy: Wide-field 1-2 GHz Research
  on Galaxy Evolution

\end{thebibliography}

\begin{appendix}
\section{Redshifts}

\begin{table*}
  \caption{\label{EMMI_POS}NTT/EMMI spectroscopic redshifts of
    SCL2243-G and I (left), SCL2243-H (middle) and SCL2243-J
    (right). Cluster members are highlighted in bold face.
    SCL2243-J is hidden behind a foreground cluster at $z=0.326$.}
  \begin{tabular}{lll|lll|lll}
    \noalign{\smallskip}
    \noalign{\smallskip}
    \hline 
    \hline 
    \noalign{\smallskip}
    \noalign{\smallskip}
    RA (J2000) & DEC (J2000) & $z$ & RA (J2000) & DEC (J2000) & $z$ & RA (J2000) & DEC (J2000) & $z$ \\ 
    \noalign{\smallskip}
    \noalign{\smallskip}
    \hline 
    \noalign{\smallskip}
    {\bf 22:41:39.99} & {\bf --09:39:11.7} & {\bf 0.433} & {\bf 22:41:42.26} & {\bf --09:38:09.1} & {\bf 0.446} & 22:40:27.68 & --09:37:27.8 & 0.314                   \\  
    {\bf 22:41:38.46} & {\bf --09:39:08.4} & {\bf 0.435} & 22:41:40.04 & --09:38:16.3 & 0.174 &                   {\bf 22:40:28.53} & {\bf --09:37:27.8} &  {\bf 0.456} \\    
    {\bf 22:41:38.13} & {\bf --09:39:08.4} & {\bf 0.433} & {\bf 22:41:38.75} & {\bf --09:38:17.9} & {\bf 0.449} & 22:40:29.10 & --09:37:27.6 & 0.317                   \\    
    22:41:36.95 & -- 09:39:15.8 & 0.622                  & {\bf 22:41:37.08} & {\bf --09:38:09.9} & {\bf 0.445} & 22:40:30.12 & --09:37:19.1 & 0.269                   \\      
    {\bf 22:41:36.57} & {\bf --09:39:09.6} & {\bf 0.439} & {\bf 22:41:35.15} & {\bf --09:38:11.4} & {\bf 0.447} & 22:40:31.03 & --09:37:35.2 & 0.251                   \\    
    {\bf 22:41:35.93} & {\bf --09:39:02.9} & {\bf 0.435} & {\bf 22:41:34.91} & {\bf --09:38:13.0} & {\bf 0.447} & 22:40:31.28 & --09:37:35.3 & 0.325                   \\    
    22:41:35.39 & --09:39:15.1 & 0.152                   & 22:41:33.22 & --09:38:02.8 & 0.423 &                   22:40:32.28 & --09:37:23.6 & 0.290                   \\      
    {\bf 22:41:34.62} & {\bf --09:39:16.9} & {\bf 0.437} & {\bf 22:41:32.89} & {\bf --09:38:01.3} & {\bf 0.453} & 22:40:33.79 & --09:37:34.7 & 0.327                   \\    
    {\bf 22:41:33.66} & {\bf --09:39:05.2} & {\bf 0.437} & {\bf 22:41:32.21} & {\bf --09:38:01.9} & {\bf 0.451} & 22:40:36.45 & --09:37:35.1 & 0.325                   \\    
    22:41:33.05 & --09:39:10.1 & 0.461                   & {\bf 22:41:31.13} & {\bf --09:38:12.9} & {\bf 0.447} & 22:40:37.54 & --09:37:32.5 & 0.709                   \\      
    22:41:32.28 & --09:39:14.0 & 0.308                   & {\bf 22:41:30.46} & {\bf --09:38:11.6} & {\bf 0.446} & 22:40:38.43 & --09:37:23.3 & 0.326                   \\      
    {\bf 22:41:31.07} & {\bf --09:39:07.8} & {\bf 0.434} & {\bf 22:41:29.70} & {\bf --09:38:10.3} & {\bf 0.446} & 22:40:39.70 & --09:37:35.8 & 0.326                   \\    
    22:41:30.58 & --09:39:11.2 & 0.396                   & 22:41:28.75 & --09:38:09.1 & 0.360 &                   22:40:40.63 & --09:37:25.3 & 0.548                   \\      
    22:41:29.98 & --09:39:11.9 & 0.217                   & 22:41:28.16 & --09:38:11.2 & 0.257 &                   22:40:41.32 & --09:37:22.9 & 0.318                   \\      
    {\bf 22:41:28.04} & {\bf --09:39:10.0} & {\bf 0.435} & 22:41:27.68 & --09:38:11.0 & 0.257 &                   22:40:42.08 & --09:37:29.3 & 0.223                   \\    
    {\bf 22:41:27.27} & {\bf --09:39:17.2} & {\bf 0.436} & {\bf 22:41:27.07} & {\bf --09:38:12.1} & {\bf 0.447} & 22:40:42.87 & --09:37:29.0 & 0.331                   \\    
    {\bf 22:41:27.03} & {\bf --09:39:11.6} & {\bf 0.435} & {\bf 22:41:25.96} & {\bf --09:38:11.5} & {\bf 0.448} & 22:40:43.67 & --09:37:28.5 & 0.222                   \\    
    {\bf 22:41:25.56} & {\bf --09:39:17.0} & {\bf 0.436} & 22:41:25.11 & --09:38:02.9 & 0.480 &                   22:40:43.89 & --09:37:32.6 & 0.327                   \\    
    22:41:25.18 & --09:39:18.1 & 0.557                   & 22:41:23.71 & --09:38:05.3 & 0.328 &                   22:40:45.16 & --09:37:28.2 & 0.201                   \\      
    22:41:24.21 & --09:39:05.4 & 0.313                   & 22:41:21.51 & --09:38:16.5 & 0.246 &                   22:40:45.81 & --09:37:29.1 & 0.323                   \\      
    22:41:21.43 & --09:39:16.3 & 0.266                   & 22:41:19.73 & --09:38:05.8 & 0.263 &                   22:40:46.41 & --09:37:26.7 & 0.324                   \\      
    22:41:20.61 & --09:39:10.7 & 0.419                   & 22:41:19.12 & --09:38:05.0 & 0.416 &                   22:40:46.80 & --09:37:33.2 & 0.326                   \\      
    {\bf 22:41:19.90} & {\bf --09:39:17.4} & {\bf 0.443} & {\bf 22:41:18.44} & {\bf --09:38:10.5} & {\bf 0.448} & 22:40:47.37 & --09:37:32.6 & 0.316                   \\    
    {\bf 22:41:19.52} & {\bf --09:39:17.4} & {\bf 0.443} & {\bf 22:41:17.55} & {\bf --09:38:11.1} & {\bf 0.449} & 22:40:47.70 & --09:37:32.6 & 0.329                   \\    
    22:41:19.23 & --09:39:13.6 & 0.333                   & 22:41:17.02 & --09:38:05.3 & 0.254 &                   22:40:48.98 & --09:37:30.7 & 0.331                   \\      
    22:41:17.48 & --09:39:02.9 & 0.175                   & 22:41:16.29 & --09:38:14.5 & 0.270 &                   22:40:49.00 & --09:37:33.1 & 0.326                   \\      
    22:41:16.41 & --09:39:09.0 & 0.554                   & 22:41:15.73 & --09:38:04.7 & 0.251 &                   22:40:50.22 & --09:37:32.5 & 0.324                   \\      
    22:41:15.25 & --09:39:11.7 & 0.554                   & 22:41:14.98 & --09:38:09.4 & 0.164 &                   {\bf 22:40:50.77}  & {\bf --09:37:21.2}  & {\bf 0.457}\\      
    {\bf 22:41:14.75} & {\bf --09:39:07.0} & {\bf 0.436} & 22:41:14.33 & --09:38:05.2 & 0.254 &                   22:40:52.93 & --09:37:29.9 & 0.326                   \\    
    {\bf 22:41:13.93} & {\bf --09:39:03.9} & {\bf 0.437} & 22:41:13.05 & --09:38:11.1 & 0.101 &                   22:40:54.50 & --09:37:27.1 & 0.325                   \\    
    {\bf 22:41:12.74} & {\bf --09:39:07.2} & {\bf 0.437} & 22:41:12.04 & --09:38:06.5 & 0.281 &                   22:40:55.52 & --09:37:26.4 & 0.324                   \\    
    {\bf 22:41:11.54} & {\bf --09:39:07.6} & {\bf 0.437} &  & &  &                                                22:40:56.16 & --09:37:25.4 & 0.327                   \\  
    {\bf 22:41:10.89} & {\bf --09:39:07.0} & {\bf 0.435} &  & &  &                                                {\bf 22:40:56.73}  & {\bf --09:37:32.3}  & {\bf 0.449}\\  
    22:41:10.53 & --09:39:01.9 & 0.190                   &  & &  &                                                22:40:57.60 & --09:37:36.7 & 0.270                   \\ 
    \hline
  \end{tabular}
  \normalsize
\end{table*}

\section{Further plots}
\begin{figure*}[t]
  \includegraphics[width=1.0\hsize]{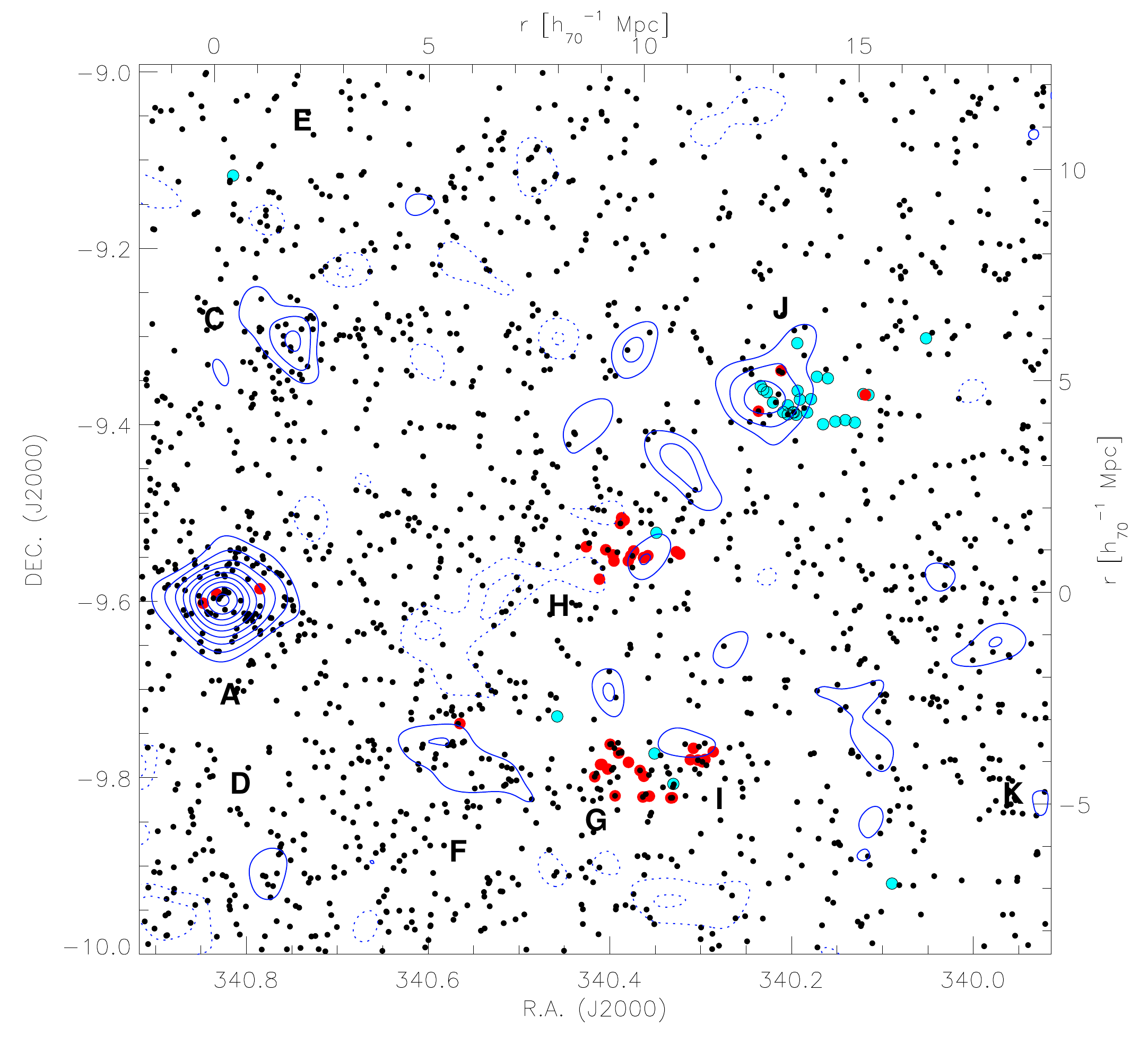}
  \caption{\label{scl2241_blue}{
      Same as Fig. \ref{scl2241_redsequ_specz_045}, but for late-type
      galaxies with $0.40\leq z_{\rm{phot}}\leq 0.52$. Note that the 
      filamentary supercluster structure is much less evident from
      these galaxies. Overlaid are galaxies with EMMI and SDSS spectra
      within $0.43<z_{\rm{spec}}<0.46$ (red dots). Blue dots with black
      circles have $0.32<z_{\rm{spec}}<0.33$. The cluster at
      RA$=340.2$ DEC$=-9.35$ is a foreground object ($z=0.326$) 
      superimposed over a sub-structure at $z=0.456$. The solid blue
      contours represent the S/N of the mass reconstruction, starting
      at $2\sigma$ and increasing in steps of $1.0\sigma$; dashed
      contours represent negative values ($2\sigma$ and $3\sigma$). 
      The smoothing length for the surface mass density was 
      $4\myarcmin0$. The upper and right axes give the physical scale
      at the main cluster redshift of $z=0.447$.}}
\end{figure*}

\section{Selected cluster images}
%\begin{figure*}[t]
%  \includegraphics[width=1.0\hsize]{scl2241_redsequ_specz_0.30.ps}
%  \caption{\label{scl2241_redsequ_specz_0.30}{Same as Fig. 
%      \ref{scl2241_redsequ_specz_0.45}, but for red sequence 
%      galaxies with photometric redshifts the range $0.27<z_p<0.37$.
%      The physical scale is given for a redshift of $z=0.326$. Note
%      that the filamentary supercluster structure has disappeared.
%      The few galaxies left at the supercluster centre are scattered
%      into lower redshift bins due to photometric redshift 
%      uncertainties $\sigma_{\Delta z}\sim0.04\,(1+z)^{-1}$.}}
%\end{figure*}

\begin{figure*}
  \includegraphics[width=1.0\hsize]{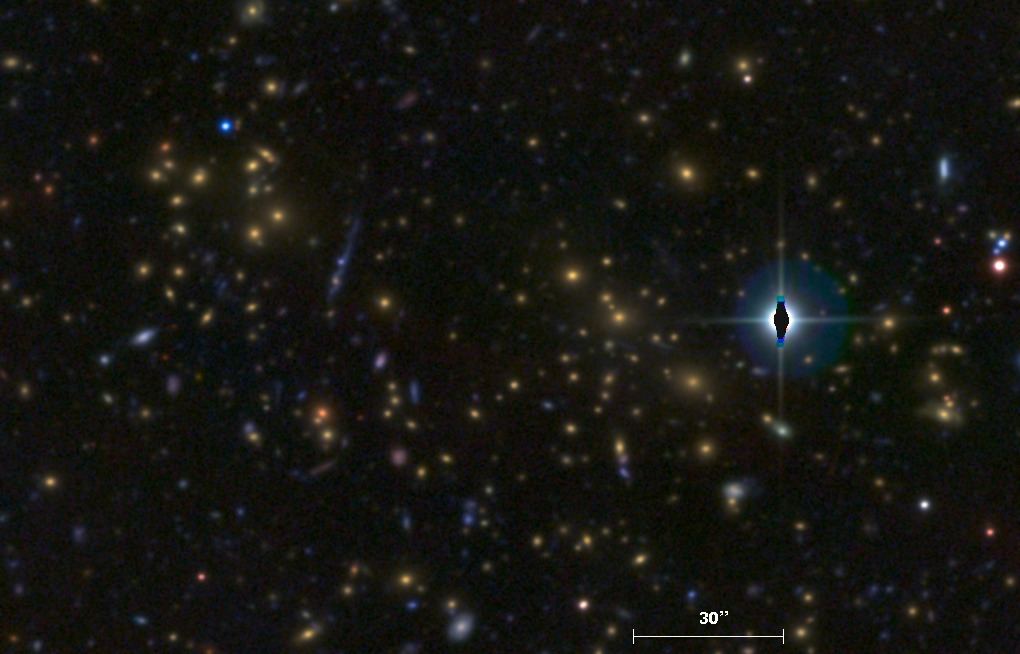}
  \caption{\label{scl2243_a}{SCL2243-A (CFHT image). Note that this
      image does not cover the full extent of this cluster. The same
      holds for all other images below, only the central or
      representative parts are shown.}}
\end{figure*}
\begin{figure*}
  \includegraphics[width=1.0\hsize]{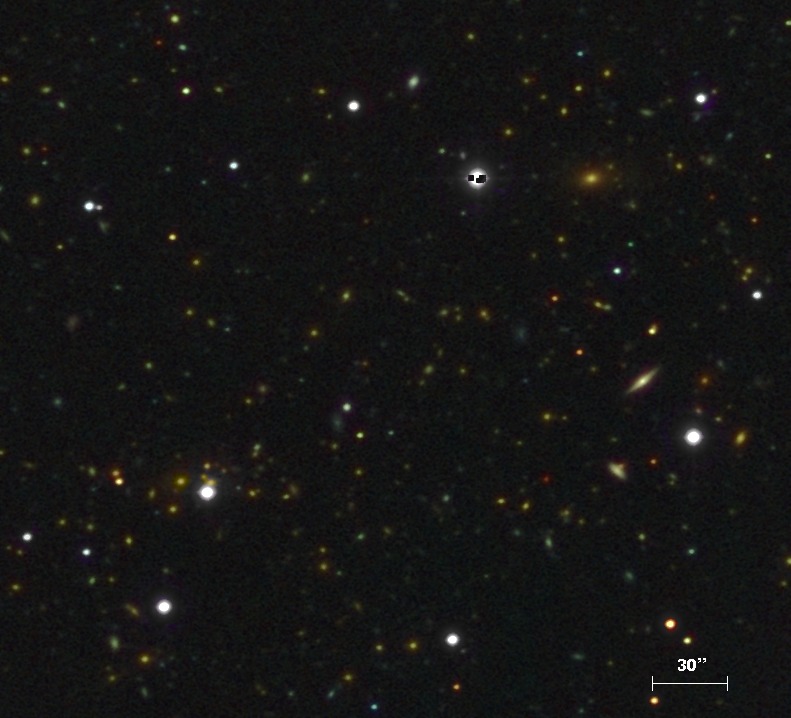}
  \caption{\label{scl2243_b}{SCL2243-B (INT image).}}
\end{figure*}
\begin{figure*}
  \includegraphics[width=1.0\hsize]{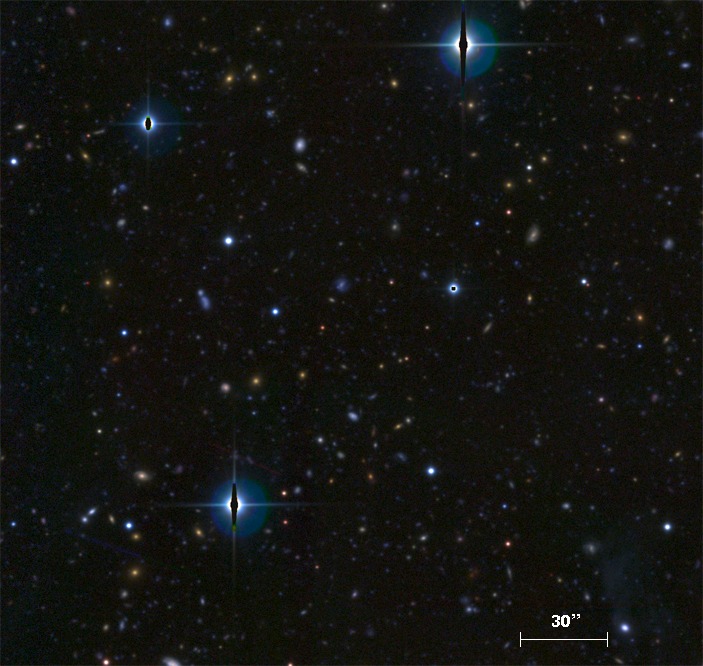}
  \caption{\label{scl2243_c}{SCL2243-C (CFHT image). Note the sparse
      distribution of early-type galaxies.}}
\end{figure*}
\begin{figure*}
  \includegraphics[width=1.0\hsize]{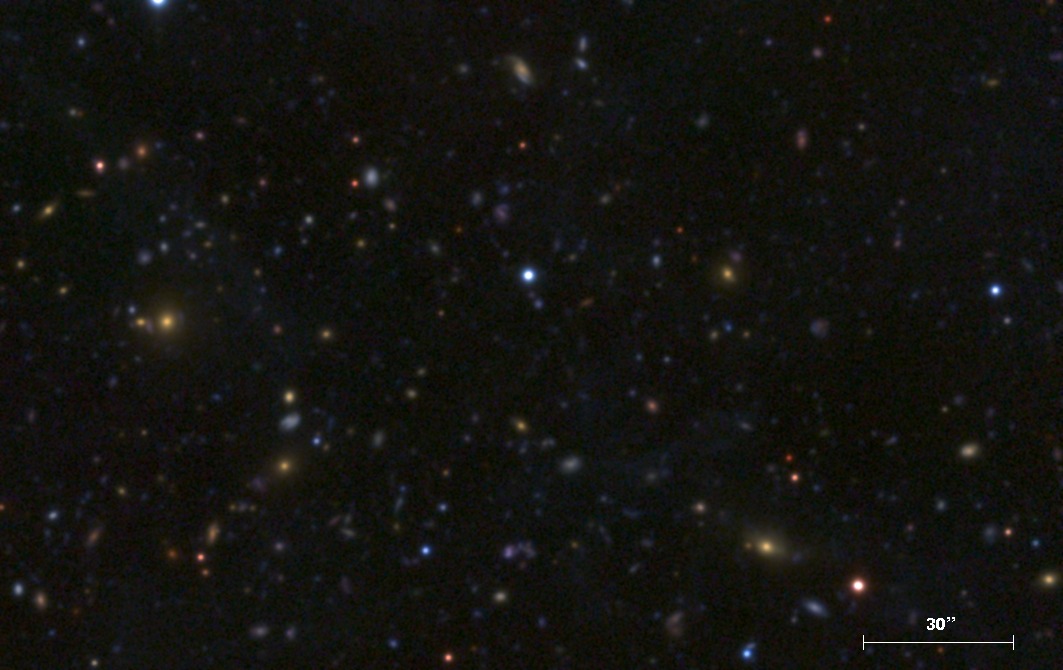}
  \caption{\label{scl2243_g}{SCL2243-G (CFHT image).}}
\end{figure*}
\begin{figure*}
  \includegraphics[width=1.0\hsize]{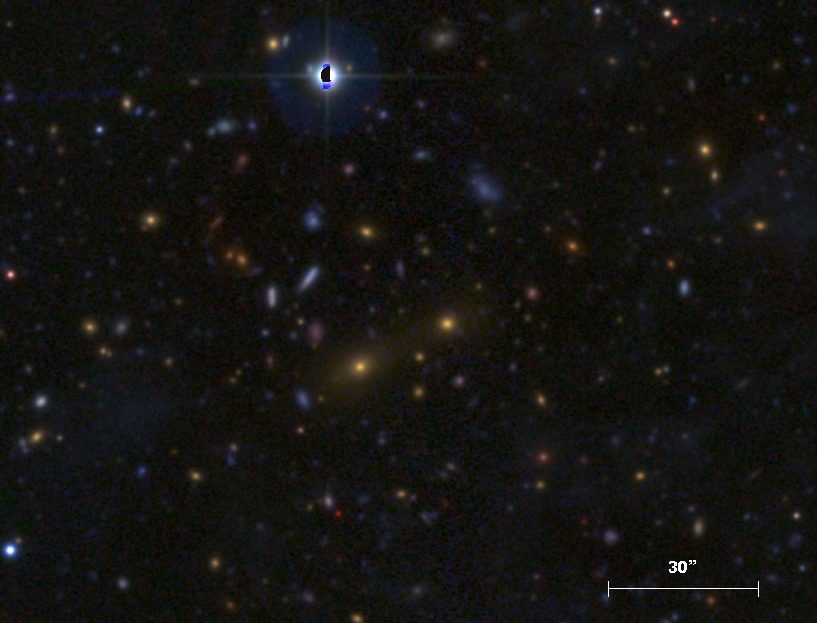}
  \caption{\label{scl2243_h}{SCL2243-H (CFHT image). The image does not
      cover the full extent of this cluster.}}
\end{figure*}
\begin{figure*}
  \includegraphics[width=0.6\hsize]{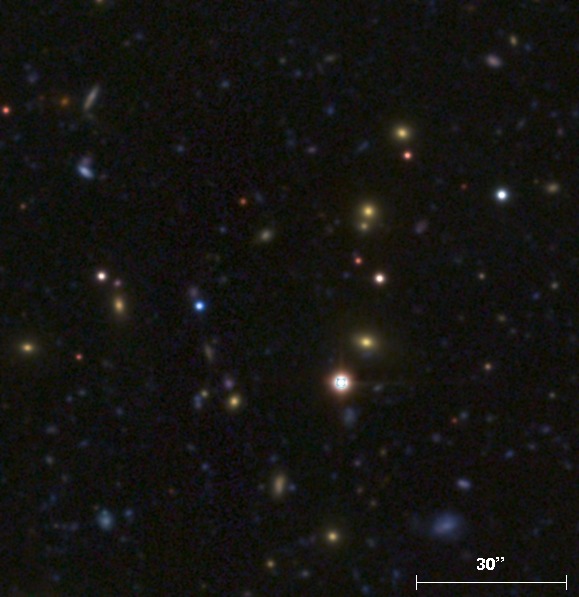}
  \caption{\label{scl2243_i}{SCL2243-I (CFHT image).}}
\end{figure*}
\begin{figure*}
  \includegraphics[width=0.6\hsize]{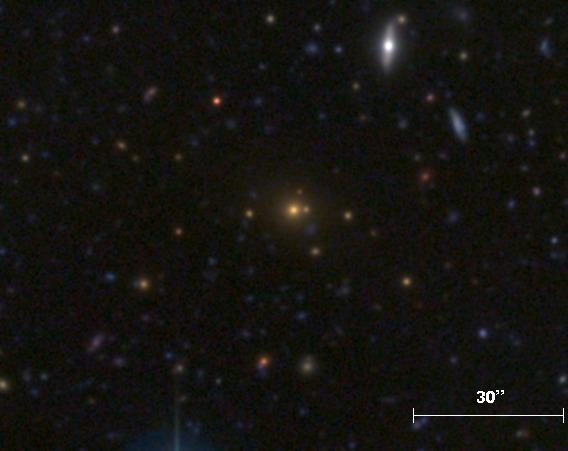}
  \caption{\label{scl2243_j}{Part of SCL2243-J (CFHT image). The
      foreground cluster at $z=0.326$ is outside the field of view.}}
\end{figure*}
\begin{figure*}
  \includegraphics[width=0.7\hsize]{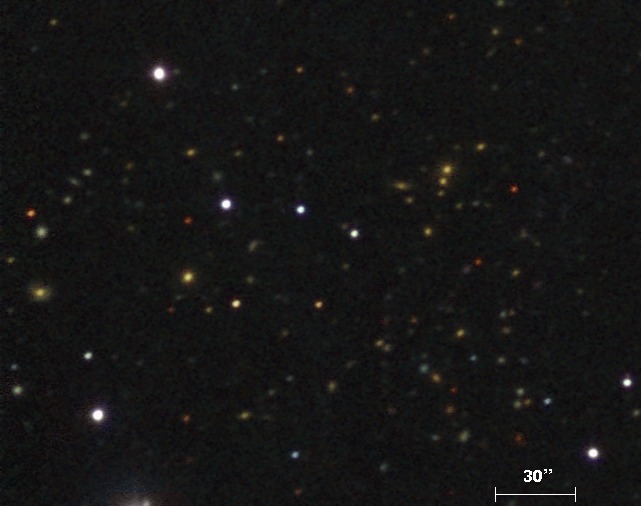}
  \caption{\label{scl2243_l}{SCL2243-L (INT image).}}
\end{figure*}
\begin{figure*}
  \includegraphics[width=1.0\hsize]{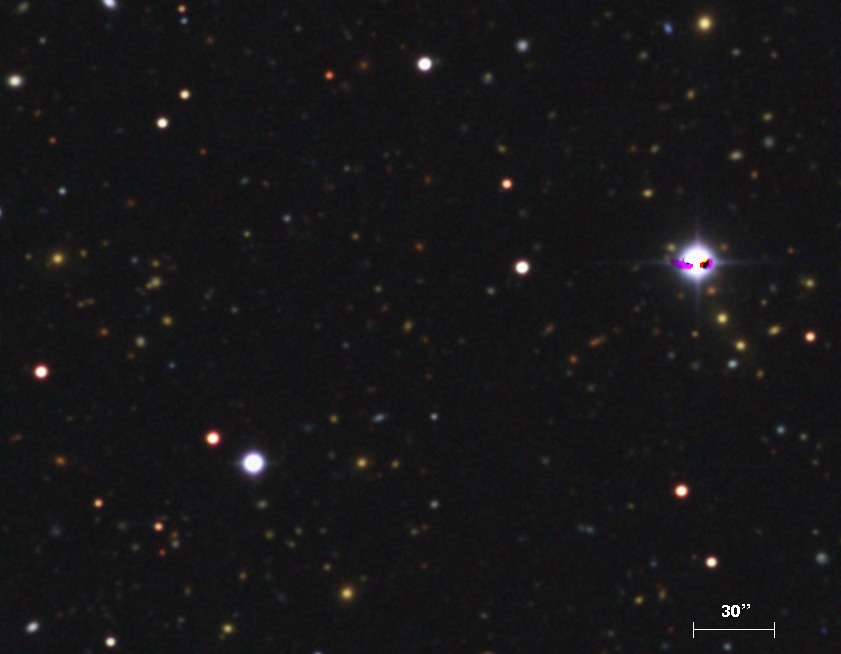}
  \caption{\label{scl2243_m}{Part of SCL2243-M (INT image).}}
\end{figure*}
\begin{figure*}
  \includegraphics[width=1.0\hsize]{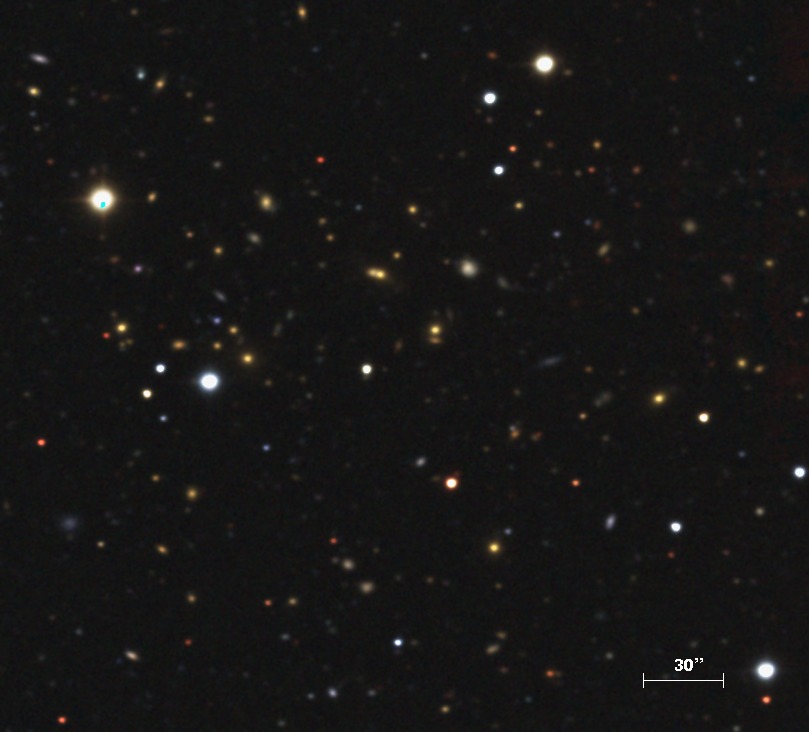}
  \caption{\label{scl2243_n}{SCL2243-N (INT image).}}
\end{figure*}

\end{appendix}
\end{document}